\documentclass[3p]{elsarticle}

\usepackage{siunitx}
\usepackage{stackengine}
\stackMath
\newcommand\tenbar[2][1]{%
  \def\useanchorwidth{T}%
  \ifnum#1>1
    \stackunder[0.5pt]{\tenbar[\numexpr#1-1\relax]{#2}}{\scriptscriptstyle\rule{1.4ex}{0.6pt}}%
  \else
    \stackunder[2pt]{#2}{\scriptscriptstyle\rule{1.4ex}{0.6pt}}%
  \fi
}

\usepackage{hyperref}
\hypersetup{
    colorlinks,
    linkcolor={magenta!80!black},
    citecolor={blue!80!black},
    urlcolor={blue!80!black}
}

\usepackage{algorithm}
\usepackage{algorithmic}

\usepackage{enumitem}
\usepackage{physics}
\usepackage{float}
\usepackage{subcaption}

\usepackage[capitalise, noabbrev]{cleveref}

\usepackage{amssymb}
\usepackage{amsmath}
\usepackage{wrapfig}
\usepackage{booktabs}
\usepackage{multirow}
\usepackage{pgffor}
\usepackage{comment}

\journal{Journal of Hydrology}

\begin{document}

\begin{frontmatter}

\title{Flood forecasting using a physics-guided conditional denoising diffusion probabilistic model}

\author[label1]{Omar Sallam\corref{cor1}}
\ead{osallam@anl.gov}

\author[label1]{David Coe}
\author[label1]{Jeremy Feinstein}
\author[label1]{Hugo Torres}
\author[label1]{Sofia Sahagun Covarrubias}
\author[label1]{Eugene Yan}

\affiliation[label1]{
    organization={Environmental Science Division, Argonne National Laboratory},
    addressline={9700 S Cass Ave},
    city={Lemont},
    postcode={60439},
    state={IL},
    country={USA}
}

\cortext[cor1]{Corresponding author}

\begin{abstract}

Flooding is the most frequent natural hazard worldwide and among the most destructive, yet high-resolution flood forecasting with full two-dimensional shallow water equation (SWE) solvers remains computationally expensive, even on modern GPU architectures. This study develops a conditional denoising diffusion probabilistic model (DDPM) that forecasts the temporal evolution of 30~m water-depth fields over a four-day horizon for the Des Plaines River watershed in northeastern Illinois. The model is trained on hydrodynamic simulations generated with SynxFlow, a multi-GPU finite-volume SWE solver, driven by 113 historical storms observed across twelve Midwestern states between 2003 and 2024, each simulated under three antecedent-wetness scenarios. To improve physical consistency and suppress generative hallucinations, the DDPM can optionally be conditioned on coarse hydrodynamic simulations at 100~m or 150~m resolution.

Because the reverse diffusion process is stochastic, the model is probabilistic by construction, and every forecast reported here is a 20-member ensemble drawn from the same trained model. Coarse-resolution conditioning reduces the four-day mean water-depth RMSE of the ensemble mean from 0.170~m without conditioning to 0.054~m and 0.048~m with 100~m and 150~m conditioning, respectively. The coarser 150~m conditioning also yields the best critical success index (0.839), false alarm ratio, CRPS and Brier score while being cheaper to run, showing that a finer conditioning resolution does not guarantee better skill. Averaging the members reduces water-depth error by 5--11\% relative to a single realization, and the ensemble spread provides spatially explicit flood exceedance probabilities. The ensembles are, however, under-dispersive, with spread-to-error ratios of 0.35--0.73: resampling the diffusion noise captures the uncertainty of the surrogate itself but is not a substitute for perturbing the meteorological forcing. Relative to a single-GPU 30~m simulation, a single-member forecast is $7.8\times$, $11.0\times$ and $16.3\times$ faster for the 100~m, 150~m and unconditioned configurations; the advantage increases with ensemble size and forecast horizon, reaching $48$--$58\times$ for a 20-member 14-day forecast. These results indicate that physics-guided diffusion models are promising probabilistic surrogates for regional-scale flood forecasting.
\end{abstract}


\end{frontmatter}

\section{Introduction}
\label{sec:introduction}

Flooding is the most frequently occurring natural and climate-related disaster worldwide \cite{CRED2025Disasters}. According to the Center for Research on the Epidemiology of Disasters (CRED), floods accounted for an average of 169 major disaster events annually over the last two decades, exceeding all other climate-related hazards, including storms, which occurred at an average rate of 108 events per year \cite{CRED2025Disasters}. During the same period, flooding was responsible for an average of 5,544 fatalities annually and caused approximately \$45.6 billion in average yearly economic losses worldwide \cite{CRED2025Disasters}. These statistics highlight the significant societal and economic burden imposed by flood hazards across both developed and developing regions.

The high frequency of flooding is closely linked to multiple hydroclimatic processes. As global temperatures rise, the atmosphere can hold more water vapor (approximately 7\% per $^\circ$C of warming, following the Clausius--Clapeyron relation), leading to more intense precipitation extremes and stronger storm systems \cite{seneviratne2021weather,rogers2025role}. In addition, tropical cyclones, atmospheric rivers, snowmelt events, and sea-level rise can all contribute directly or indirectly to flooding. Consequently, flooding often represents the final and most widespread manifestation of a range of climate-related hazards, including pluvial flooding from intense rainfall, fluvial flooding resulting from river overflow, flash flooding in rapidly responding catchments, and coastal flooding driven by storm surge and sea-level rise \cite{chen2026global}.

Flood events can result in loss of life, injuries, displacement of communities, property damage, transportation disruption, power and water outages, supply-chain interruption, soil erosion, water contamination, ecosystem degradation, and public-health risks such as disease outbreaks, mold growth, microbial contamination, and toxic chemical exposure \cite{ipcc2022wg2}. Therefore, reliable operational flood forecasting has become a critical component of disaster risk reduction and climate adaptation strategies. Accurate medium- and long-range flood forecasts support evacuation planning, identification of high-risk areas, protection of critical infrastructure and energy-generation facilities, and continuity of essential public services \cite{wmo2022earlywarnings}.

Operational flood forecasting systems commonly integrate meteorological forecasts, hydrologic rainfall--runoff models, and hydraulic inundation models \cite{wmo_floodforecasting}. For urban and regional-scale flood forecasting applications, including pluvial, fluvial, and flash flooding, shallow water equation (SWE)-based numerical models are widely used. These models may be implemented as one-dimensional river-routing models, two-dimensional inundation models, or simplified hydraulic formulations designed to reduce computational cost \cite{teng2017flood}.

Among these approaches, full two-dimensional SWE models provide a high level of physical fidelity because they explicitly resolve spatially varying water depths, flow velocities, flood-wave propagation, wetting and drying fronts, backwater effects, and floodplain connectivity \cite{teng2017flood}. However, these models remain computationally expensive for operational forecasting applications, particularly for large watersheds, regional- and state-scale domains, high-resolution inundation mapping at 30~m or finer spatial resolutions, and ensemble forecasting frameworks designed to quantify uncertainty in storm trajectory, precipitation intensity, antecedent watershed conditions, and boundary conditions \cite{cloke2009ensemble}.

To bridge the gap between high-fidelity flood forecasting and the computational limitations associated with large-scale numerical simulations, as well as the scarcity of in situ observations, machine learning has emerged as a promising surrogate modeling approach for computationally efficient, large-scale, and high-resolution flood forecasting. By learning the relationships between hydroclimatic forcings, watershed characteristics, and flood responses from historical data and numerical simulations, these models can provide rapid predictions suitable for operational forecasting and early warning applications \cite{asif2025review,liu2025comprehensive}.

A wide range of machine learning approaches has been investigated for flood forecasting. These include tree-based methods such as Random Forest \cite{zahura2020training} and XGBoost \cite{ma2021xgboost,lee2024predicting}; physics-guided models, including Physics-Informed Neural Networks (PINNs) \cite{liu2025pinn}, Fourier Neural Operators (FNOs) \cite{sun2023rapid,xu2024large,xu2025floodcastbench,kumar2025temporal}, and Conditional Latent Dynamics Networks (CLDNet) \cite{si2026toward}; graph-based approaches such as Graph Neural Networks (GNNs) \cite{bentivoglio2023rapid,kazadi2024floodgnn,kazadi2024pluvial,bentivoglio2025multi,taghizadeh2025interpretable,acosta2025physics}; convolutional vision-based architectures including Convolutional Neural Networks (CNNs) \cite{guo2022data,pradhan2023spatial}, Residual Networks \cite{choi2025flo}, and U-Net models \cite{lowe2021u,wang2024time,herath2025subgrid,he2023deep}; sequence-learning models such as Long Short-Term Memory (LSTM) networks \cite{zhou2021rapid,nearing2024global,cohen2026extending}; and generative probabilistic approaches including Generative Adversarial Networks (GANs) \cite{hofmann2021floodgan,do2023generalizing,huang2025high,yin2024fast,shen2026attention}, Denoising Diffusion Probabilistic Models (DDPMs) \cite{shao2024data,ou2025probabilistic,neo2026flood}, and hybrid LSTM-vision-generative architectures \cite{yang2025diffusion,chen2023fast,yang2024rapid,liao2025accelerating,wang2026hybrid}.

Tree-based machine learning methods are a widely used class of predictive models owing to their ability to capture complex nonlinear relationships, handle heterogeneous input features, and provide robust performance with relatively limited feature engineering. Random Forest \cite{breiman2001random} employs an ensemble of independently trained decision trees, whereas Gradient Boosting and its widely adopted implementation XGBoost \cite{friedman2001greedy,chen2016xgboost} construct trees sequentially to iteratively reduce prediction errors. These methods have been successfully applied to flood forecasting problems such as street-level coastal flooding \cite{zahura2020training} and peak flood-depth prediction \cite{lee2024predicting}. However, existing applications primarily focus on point-based, street-segment, or peak-value predictions rather than continuous spatiotemporal inundation forecasting. Because tree-based models operate on tabular features and do not explicitly learn spatial or temporal dependencies, their scalability to large regional domains and high-resolution two-dimensional flood forecasting remains limited.

Physics-guided machine learning models seek to improve predictive accuracy and physical consistency by incorporating governing equations, conservation laws, or other forms of scientific knowledge into the learning process. Among the most prominent approaches are Physics-Informed Neural Networks (PINNs) \cite{raissi2019physics} and Fourier Neural Operators (FNOs) \cite{li2020fourier}. PINNs directly incorporate partial differential equation residuals into the loss function, enabling physically consistent predictions while reducing reliance on large training datasets. In contrast, FNOs learn mappings between function spaces through spectral convolutions in the Fourier domain, enabling efficient approximation of complex solution operators. While PINNs offer strong physical interpretability, they often require expensive optimization procedures and long training times, particularly for large computational domains, and may struggle with highly nonlinear flow regimes and unseen hydrologic conditions. FNOs generally provide significantly faster inference and improved scalability; however, they require large volumes of training data generated from numerical simulations and may exhibit degraded performance when applied to unseen events or boundary conditions \cite{sun2023rapid,xu2024large,xu2025floodcastbench,kumar2025temporal}.

Graph Neural Networks (GNNs) are designed to learn from graph-structured data by representing entities as nodes and interactions as edges \cite{scarselli2008graph,kipf2016semi,battaglia2018relational}. Through message-passing operations, GNNs effectively capture connectivity patterns and spatial dependencies, making them particularly attractive for representing river networks and watershed systems. Consequently, GNNs have been increasingly applied to flood forecasting and hydrologic prediction problems \cite{bentivoglio2023rapid,kazadi2024floodgnn,kazadi2024pluvial,bentivoglio2025multi,taghizadeh2025interpretable,acosta2025physics}. However, training costs and memory requirements increase substantially as graph size and spatial resolution grow. Furthermore, representing high-resolution two-dimensional inundation fields often requires extremely large graph structures, limiting scalability for regional-scale flood forecasting applications.

Long Short-Term Memory (LSTM) networks are a specialized class of recurrent neural networks developed to capture long-term temporal dependencies in sequential data while mitigating the vanishing gradient problem \cite{elman1990finding,hochreiter1997long}. Through memory cells and gating mechanisms, LSTMs effectively learn relationships between past and future states, making them highly suitable for time-series forecasting applications. As a result, LSTM-based models have been widely applied to streamflow forecasting, water-level prediction, and hydrologic time-series analysis \cite{zhou2021rapid,nearing2024global,cohen2026extending}. Nevertheless, LSTMs primarily learn temporal relationships and do not explicitly represent spatial flood dynamics. Consequently, they are generally better suited for point-scale forecasting at gauging stations than for generating high-resolution two-dimensional inundation maps over large geographic domains.

Convolutional Neural Networks (CNNs) are designed to process spatially structured data through convolutional operations that automatically learn hierarchical feature representations. A widely adopted CNN architecture is the U-Net \cite{ronneberger2015u}, which combines encoder-decoder pathways with skip connections to preserve fine-scale spatial information during reconstruction. CNNs and U-Net architectures have demonstrated promising performance for flood inundation mapping, surrogate hydraulic modeling, and spatial prediction tasks \cite{lowe2021u,wang2024time,herath2025subgrid,he2023deep}. However, conventional convolution operations are inherently local and may struggle to capture long-range dependencies across large watersheds and floodplains.

Generative Adversarial Networks (GANs) have emerged as a powerful class of generative models capable of producing realistic synthetic data through an adversarial training process between a generator and discriminator \cite{goodfellow2014generative}. Owing to their ability to generate high-resolution spatial fields, GANs have been applied to flood inundation mapping, surrogate hydraulic modeling, and super-resolution tasks \cite{hofmann2021floodgan,do2023generalizing,huang2025high,yin2024fast,shen2026attention}. However, GAN training is often unstable and highly sensitive to hyperparameter selection. Furthermore, GANs are susceptible to mode collapse \cite{salimans2016improved,metz2016unrolled,arjovsky2017wasserstein}, where the model generates only a subset of plausible outcomes rather than fully representing the underlying flood distribution. This limitation is particularly problematic for probabilistic flood forecasting and uncertainty quantification.

Denoising Diffusion Probabilistic Models (DDPMs) have recently emerged as a powerful alternative generative framework capable of producing high-fidelity spatial fields through an iterative denoising process \cite{ho2020denoising}. Unlike GANs, diffusion models are trained with a stable denoising objective derived from a variational bound on the data likelihood and are substantially less susceptible to mode collapse. Recent studies have demonstrated the potential of diffusion models for probabilistic flood forecasting and spatiotemporal prediction \cite{shao2024data,ou2025probabilistic,neo2026flood,yang2025diffusion}. Nevertheless, the integration of high-resolution flood inundation forecasting, large-scale spatial coverage, and temporal evolution within a single diffusion-based framework remains largely unexplored.

Although DDPMs generally require higher inference costs due to the iterative denoising process, recent advances in latent diffusion models \cite{rombach2022high}, accelerated samplers, and consistency-based diffusion methods \cite{song2023consistency} have substantially reduced this computational burden. These characteristics make diffusion-based models particularly attractive for next-generation flood forecasting systems that require high-resolution spatial prediction, probabilistic forecasting, uncertainty quantification, and regional-scale spatiotemporal scalability.

To mitigate the risk of generative hallucinations and improve physical consistency in flood forecasting applications, previous studies have explored conditioning generative models on the outputs of computationally inexpensive physics-based simulations. These conditioning inputs may originate from coarse-resolution or lower-fidelity flood modeling frameworks that provide physically plausible large-scale flow structures while avoiding the computational cost of fully resolved, high-resolution two-dimensional SWE solutions. The generative model then learns to reconstruct the higher-resolution, higher-frequency, or higher-fidelity inundation dynamics conditioned on these physically informed inputs, thereby combining computational efficiency with enhanced predictive fidelity. Examples include temporal super-resolution using FNOs \cite{kumar2025temporal} and spatial super-resolution using U-Net \cite{he2023deep,herath2025subgrid}, ResNet \cite{choi2025flo}, and GAN \cite{yin2024fast,shen2026attention} architectures.

This paradigm closely resembles the super-resolution framework that has been extensively investigated in turbulent flows and meteorological modeling, where machine learning models reconstruct high-resolution flow fields from coarse-resolution or mesoscale simulations while preserving the underlying low-frequency, large-scale physical structures and dynamics \cite{fukami2019super,fukami2021machine,buster2025second,benton2025super,stengel2020adversarial,sallam2025tropical}.

In this study, we develop a conditional DDPM for two-dimensional spatiotemporal flood inundation forecasting. The model learns the conditional probability distribution of flood dynamics and predicts the temporal evolution of high-resolution water depth, rather than a binary inundation extent or a single peak-flood snapshot.

To improve physical consistency and reduce the risk of generative hallucinations, the diffusion model can optionally be conditioned on low-resolution hydrodynamic simulations generated by the GPU-accelerated SynxFlow shallow-water-equation solver. This combines the computational efficiency of coarse-resolution numerical simulations with the ability of diffusion models to reconstruct high-resolution inundation patterns.

A further consequence of the diffusion formulation, and one that is central to this work, is that the model is probabilistic by construction rather than by extension. Ensemble prediction is the standard operational route to quantifying forecast uncertainty in both meteorology \cite{leutbecher2008ensemble} and flood forecasting \cite{cloke2009ensemble}, and diffusion models have recently been shown to produce skillful, well-calibrated ensembles for global weather prediction \cite{price2025probabilistic}. Because the reverse denoising process is a stochastic Markov chain, repeated evaluation of the trained network under identical conditioning returns independent draws from the learned conditional distribution of flood states. Forecast ensembles can therefore be generated without retraining, without perturbing the inputs, and without any architectural modification, and the resulting spread provides spatially explicit probabilistic flood information. We exploit this property explicitly, generating multi-member ensembles for the withheld testing events and verifying them with standard probabilistic scores. This distinguishes the present framework from deterministic surrogate models, for which uncertainty quantification requires either input perturbation, multi-model construction, or auxiliary variance estimation.

The proposed model is trained using two decades of historical precipitation events (2003--2024) observed across the Midwestern United States. In contrast to approaches relying on synthetic rainfall forcing, the training dataset is constructed from observed two-dimensional rainfall fields that preserve the spatial and temporal variability of precipitation patterns observed in nature. The corresponding target inundation fields are generated using the validated SynxFlow two-dimensional shallow-water-equation solver.

The framework is evaluated for the Des Plaines River watershed at a spatial resolution of 30~m, corresponding to $4{,}188{,}832$ active computational cells. A total of 113 historical precipitation events, each simulated under three different antecedent hydrologic conditions, are used for model development and evaluation, giving 339 hydrodynamic simulations per grid resolution. Six events are withheld entirely from training and used for evaluation. The model is trained on patches of $15.36 \times 15.36~\mathrm{km}^2$ rather than on the entire watershed domain. This patch-based formulation enables efficient training on large datasets and, although only a single watershed is examined here, is intended to ease the later incorporation of additional watersheds and the extension toward regional-scale forecasting.

The main contributions of this work are fourfold. First, we develop a conditional DDPM that forecasts the full four-day evolution of 30~m water depth in a single network evaluation, trained on shallow-water simulations of observed rather than synthetic storms. Second, we compare 100~m, 150~m and no hydrodynamic conditioning in a controlled setting and show that the coarser conditioning can be both more accurate and cheaper. Third, we evaluate the native diffusion ensembles probabilistically, using proper scoring rules, spread--skill analysis, rank histograms and reliability diagrams, and identify their under-dispersion. Finally, we compare the cost of the surrogate with that of the hydrodynamic solver on the same single GPU, including the dependence of the speedup on ensemble size.

The remainder of the paper is organized as follows. Section~\ref{sec:dataset} describes the study area, the SynxFlow solver and its validation, and the generation of the training dataset. Section~\ref{sec:methodology} presents the conditional DDPM, its conditioning variables, the ensemble construction and the training procedure. Section~\ref{sec:results} reports the deterministic and probabilistic skill of the forecasts and their computational cost, and Section~\ref{sec:conclusion} summarizes the conclusions and outlines future work.

\clearpage
\newpage
\section{Dataset generation}
\label{sec:dataset}

\subsection{Study area}

\begin{figure}[h]
    \centering

    \begin{subfigure}{0.24\linewidth}
        \centering
        \includegraphics[width=\linewidth]{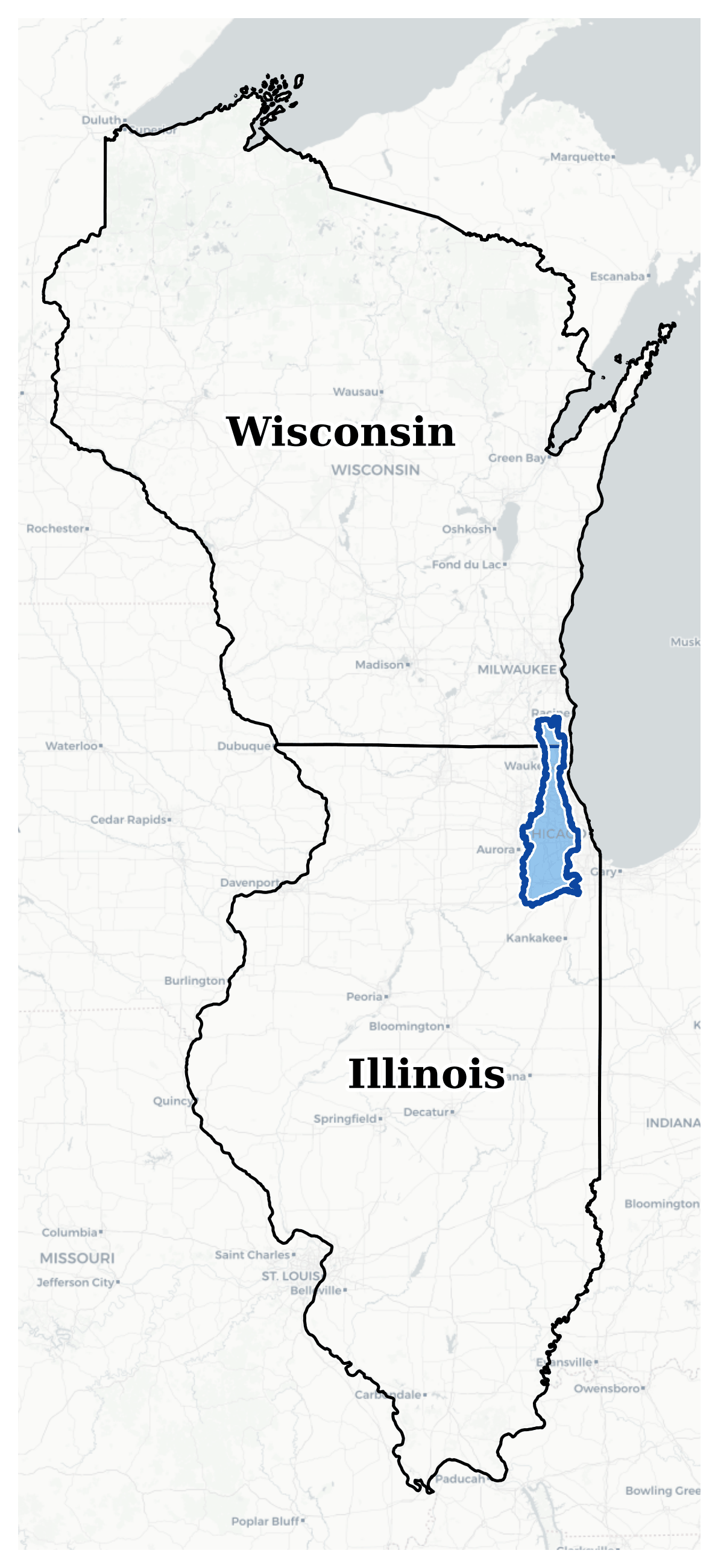}
        \caption{Des Plaines river watershed location}
    \end{subfigure}
    \hfill
    \begin{subfigure}{0.75\linewidth}
        \centering
        \includegraphics[width=\linewidth]{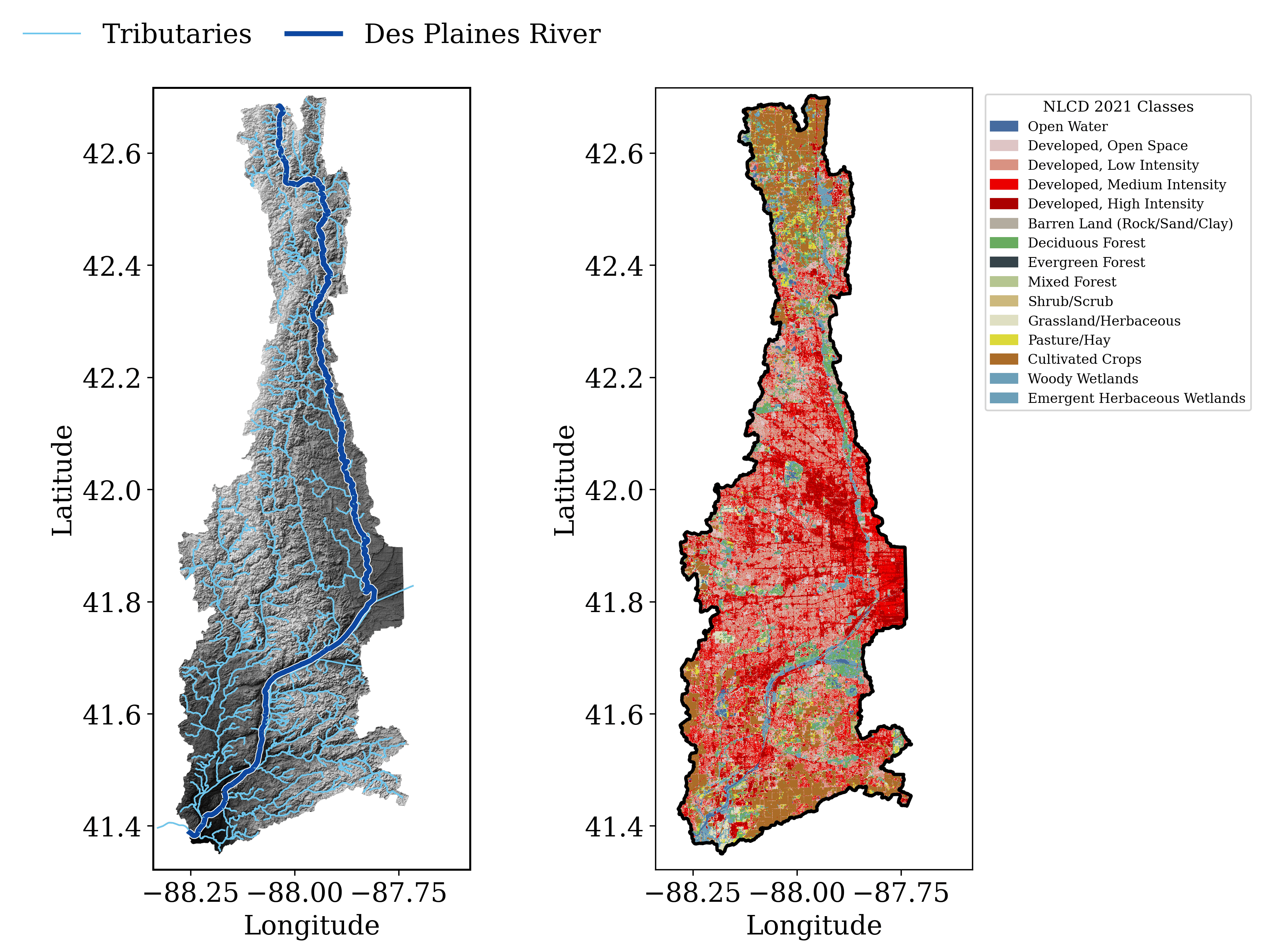}
        \caption{Main river and tributary network overlaid on the watershed terrain elevation (left), and land cover distribution derived from the NLCD 2021 dataset (right).}
    \end{subfigure}

    \caption{
    Geographic location of the Des Plaines River watershed and corresponding terrain, hydrologic network, and land cover.
    }

    \label{fig:study_area}
\end{figure}

In this study, the Des Plaines River watershed, corresponding to Hydrologic Unit Code (HUC-8) 07120004, was selected as the study area. The watershed is located in northeastern Illinois near the greater metropolitan Chicago area, with a small upstream portion extending into southern Wisconsin, as shown in Figure~\ref{fig:study_area}a. The watershed represents one of the major river systems in the urbanized Chicago region and has experienced multiple historical flooding events associated with extreme precipitation and urban runoff conditions.

The total drainage area of the watershed is approximately 3,770~km$^2$, the mean annual river discharge is on the order of 108~m$^3$/s (3,800~ft$^3$/s), and the river flows from north to south through the watershed domain, as shown in Figure~\ref{fig:study_area}b.

The topographic information used in this study was derived from the United States Geological Survey (USGS) 3D Elevation Program (3DEP) digital elevation model (DEM) at a spatial resolution of 30~m. The river network was obtained from the USGS National Hydrography Dataset (NHD), as shown in Figure~\ref{fig:study_area}b. Land use and land cover information were obtained from the National Land Cover Database (NLCD) 2021 dataset at 30~m spatial resolution, as illustrated in Figure~\ref{fig:study_area}b.

All datasets were projected onto the USA Contiguous Albers Equal Area Conic projection (EPSG:5070). At the adopted 30~m computational grid resolution the full raster domain comprises $5{,}075 \times 1{,}661$ cells, of which $4{,}188{,}832$ fall inside the watershed boundary and constitute the active computational cells. The corresponding coarse-resolution domains used for conditioning contain $1{,}523 \times 499$ cells at 100~m and $1{,}015 \times 333$ cells at 150~m.

As shown in Figure~\ref{fig:study_area}b, the watershed contains heterogeneous land use distributions. The central portion of the watershed is dominated by medium- to high-density developed urban areas associated with the Chicago metropolitan region. In contrast, the northern and southern regions are characterized by lower urbanization levels, with land cover primarily consisting of cultivated croplands, vegetation, and open green areas.

\subsection{Numerical model}

The flood inundation datasets used in this study were generated using SynxFlow (formerly HiPIMS), a high-performance multi-GPU hydrodynamic modeling framework designed for large-scale flood simulations based on the two-dimensional shallow water equations (SWEs) \cite{xia2019full, xia2025synxflow}. SynxFlow employs a Godunov-type finite volume scheme to solve the conservative form of the depth-averaged SWEs over structured Cartesian grids, enabling efficient and scalable modeling for high-resolution watershed-scale applications.

The governing equations can be written as:

\begin{equation}
\label{eq:swe_original}
    \partial_t q + \nabla \cdot \mathbf{F}(q) = R + S_b + S_f,
    \qquad (t,(x,y)) \in [0,T] \times \Omega,
\end{equation}

where $q$ denotes the vector of conserved variables, $\mathbf{F}(q)$ represents the flux tensor, and $R$, $S_b$, and $S_f$ correspond to the rainfall/infiltration, bed slope, and friction source terms, respectively. These quantities are defined as:

\begin{align*}
    q &=
    \begin{bmatrix}
        h \\
        hu \\
        hv
    \end{bmatrix},
    \qquad
    \mathbf{F} =
    \begin{bmatrix}
        uh & vh \\
        u^2h + \tfrac{1}{2}gh^2 & uvh \\
        uvh & v^2h + \tfrac{1}{2}gh^2
    \end{bmatrix},
    \\
    R &=
    \begin{bmatrix}
        r - I \\
        0 \\
        0
    \end{bmatrix},
    \qquad
    S_b =
    \begin{bmatrix}
        0 \\
        -gh\partial_x b \\
        -gh\partial_y b
    \end{bmatrix},
    \qquad
    S_f =
    \begin{bmatrix}
        0 \\
        -C_f u \sqrt{u^2+v^2} \\
        -C_f v \sqrt{u^2+v^2}
    \end{bmatrix}.
\end{align*}

Here, $h$ denotes the water depth, $u$ and $v$ are the depth-averaged velocity components in the $x$ and $y$ directions, respectively, and $b$ represents the bed elevation. The Manning-based friction coefficient is given by:

\begin{equation}
    C_f = g n^2 h^{-1/3},
\end{equation}

where $g$ is the gravitational acceleration and $n$ is Manning's roughness coefficient. A spatially variable roughness field was prescribed from the NLCD land cover classes, using $n = 0.02$~s/m$^{1/3}$ over open-water classes and $n = 0.05$~s/m$^{1/3}$ over land classes.

To ensure numerical stability and robust shock-capturing capabilities in the presence of rapidly varying topography and wetting-drying fronts, SynxFlow utilizes the Surface Reconstruction Method (SRM) proposed in \cite{xia2017efficient} for the discretization of both flux and bed-slope source terms. In addition, the stiff friction source terms are treated using a fully implicit formulation \cite{xia2018new}, improving numerical stability for high-resolution urban flood simulations. The notation associated with the governing equations is summarized in Table~\ref{tab:swe_notation}.

\begin{table}[!htb]
\centering
\caption{Notation for the shallow water equations in Eq.~\eqref{eq:swe_original}.}
\label{tab:swe_notation}
\begin{tabular}{ll}
\toprule
\textbf{Symbol} & \textbf{Description} \\
\midrule
$h$ & Water depth \\
$u, v$ & Depth-averaged velocity components in the $x$ and $y$ directions \\
$hu, hv$ & Unit-width discharges in the $x$ and $y$ directions \\
$g$ & Gravitational acceleration ($9.81~\mathrm{m/s^2}$) \\
$b$ & Bed elevation \\
$n$ & Manning roughness coefficient \\
$C_f$ & Friction coefficient ($g n^2 h^{-1/3}$) \\
$r, I$ & Rainfall and infiltration rates \\
$q$ & Conserved state vector \\
$\mathbf{F}$ & Flux tensor \\
$S_b$ & Bed slope source term \\
$S_f$ & Friction source term \\
\bottomrule
\end{tabular}
\end{table}

\subsection{Validation}

The hydrodynamic model for the Des Plaines River watershed was validated using the historical flood event that occurred across northern and central Illinois during April 17--18, 2013. This event was one of the most severe floods experienced in the Chicago metropolitan region over the past several decades and has been extensively documented in previous studies \cite{campos2015numerical}.

The April 2013 storm system produced extreme precipitation accumulations over large portions of Illinois and neighboring states. On April 18, 2013, a state of emergency was declared in Illinois, and by April 21, 44 of the state's 102 counties had been designated as disaster areas. According to National Weather Service (NWS) observations, the maximum 24-hour accumulated rainfall reached approximately 201.4~mm, 171.1~mm, and 136.4~mm across Illinois, Iowa, and Missouri, respectively \cite{campos2015numerical}.

The flood event resulted in four fatalities and caused direct economic losses exceeding \$465 million, while thousands of residential and commercial properties reported flood damages \cite{nws2013floodloss}. Due to its severe hydrometeorological impacts and the availability of observational data, the April 2013 event provides a suitable benchmark for validating large-scale flood inundation simulations within the Des Plaines River watershed.

The validation simulations covered the period from April 15 to April 23, 2013, allowing the model to capture both the rising and recession limbs of the flood hydrographs. Precipitation forcing was obtained from the NCEP/NWS Stage IV multisensor (radar and rain gauge) precipitation analysis with a spatial resolution of approximately $4 \times 4$~km$^2$.

\begin{figure*}[t]
    \centering
    \captionsetup[subfigure]{font=scriptsize}

    \newcommand{\figw}{0.99\linewidth}

    \begin{subfigure}{\figw}
        \centering
        \includegraphics[width=\linewidth]
        {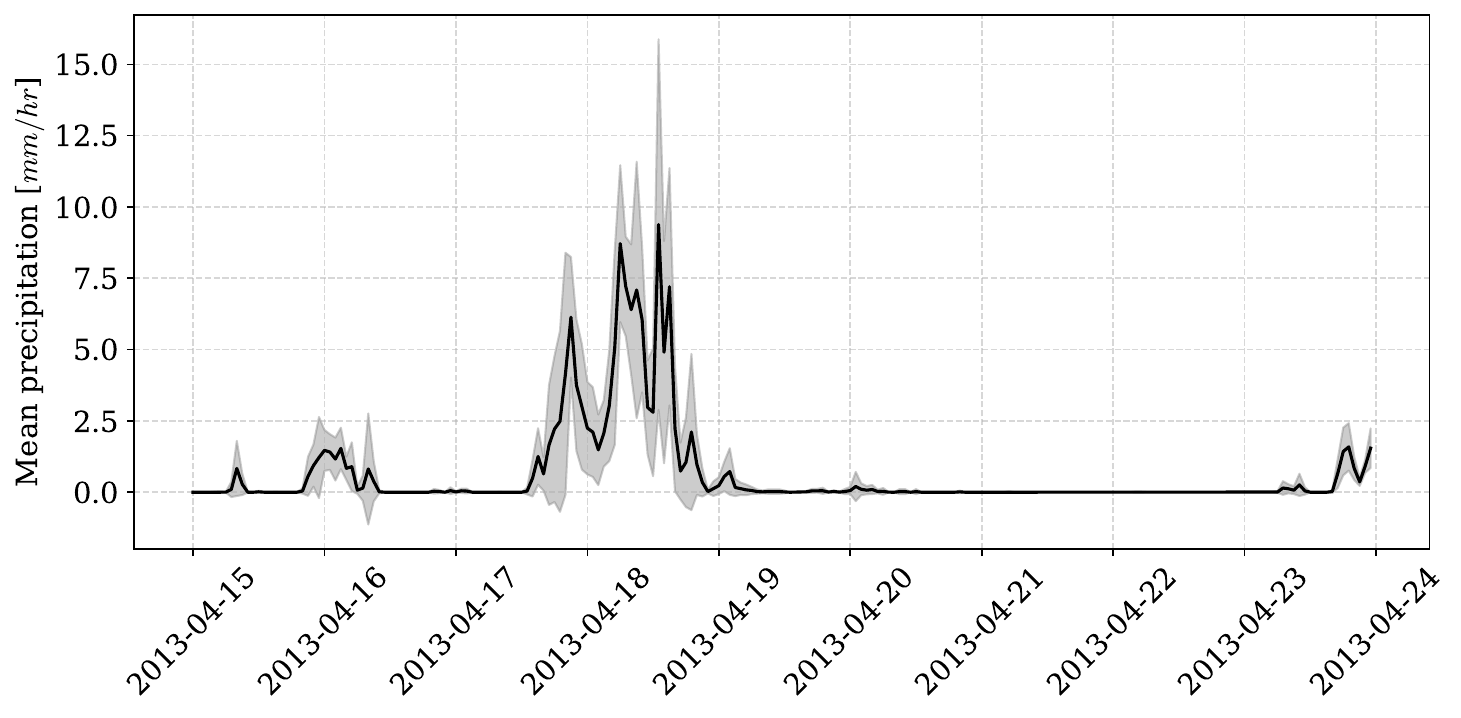}
        \caption{Temporal evolution of the spatially averaged precipitation intensity for the April 2013 Chicago flood historical event.}
    \end{subfigure}

    \vspace{0.4em}

    \begin{subfigure}{\figw}
        \centering
        \includegraphics[width=0.99\linewidth]
        {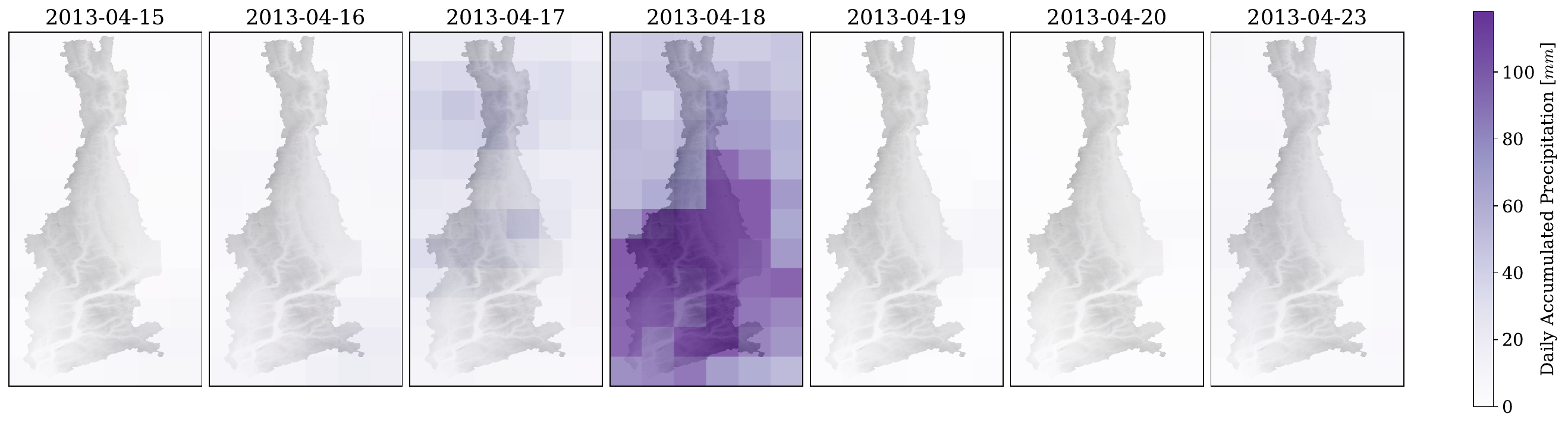}
        \caption{Daily accumulated precipitation fields overlaid on the DEM for the April 2013 Chicago flood historical event.}
    \end{subfigure}

    \caption{
    Spatiotemporal characteristics of the April 2013 Chicago flood precipitation historical event used in the numerical simulations.
    }

    \label{fig:precipitation-2013}
\end{figure*}

Figure~\ref{fig:precipitation-2013}a presents the spatially averaged precipitation time series over the Des Plaines River watershed during the April 2013 flood event, while Figure~\ref{fig:precipitation-2013}b illustrates the corresponding spatial distribution of the daily accumulated precipitation obtained from the Stage IV radar product. The precipitation contours show that the majority of the rainfall accumulation occurred on April 18, 2013, which corresponds to the peak precipitation period associated with the historical flood event.

\begin{figure}
    \centering
    \includegraphics[width=1\linewidth]{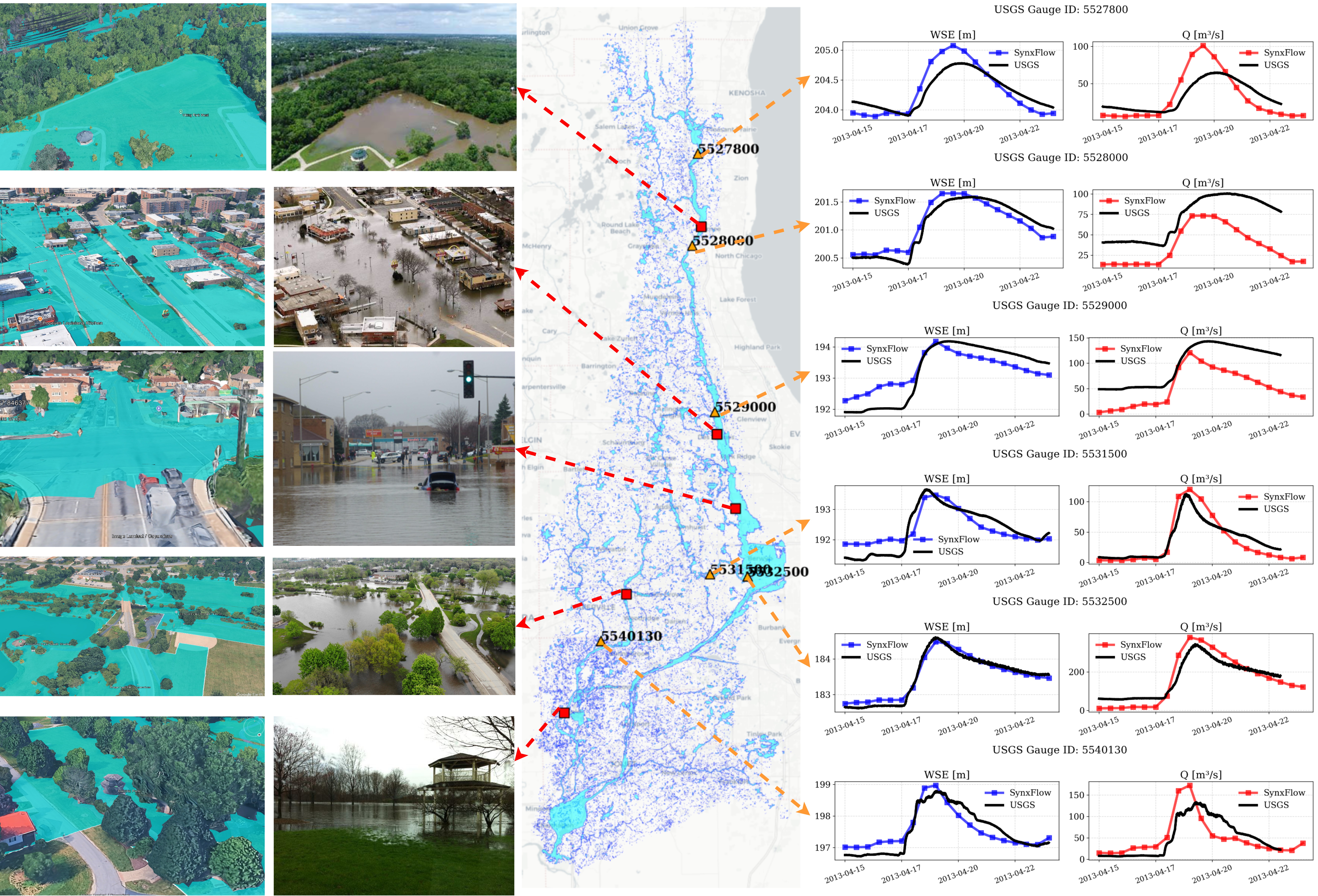}
    \caption{Validation of the SynxFlow hydrodynamic model for the Des Plaines River watershed during the April 2013 Chicago historical flood event. Flood inundation extent together with the locations of the USGS stream gauges and available community social media photographs used for validation (middle). Three-dimensional visualization of the simulated flood inundation polygons in Google Earth compared with corresponding community photographs collected during the flood event (left). Comparison between simulated and observed water surface elevation and discharge hydrographs at selected USGS gauge locations (right).}
    \label{fig:validation}
\end{figure}

For quantitative validation, six United States Geological Survey (USGS) stream gauges located within the Des Plaines River basin were selected, as shown in Figure~\ref{fig:validation}. These include four gauges located along the main stream of the Des Plaines River: USGS 05527800 (at Russell), USGS 05528000 (near Gurnee), USGS 05529000 (near Des Plaines), and USGS 05532500 (at Riverside). In addition, two tributary gauges were considered: USGS 05531500 (Salt Creek at Western Springs) and USGS 05540130 (West Branch DuPage River near Naperville). These gauges were selected because the associated river channels are sufficiently resolved at the adopted 30~m DEM resolution.

At each gauge location, a cross-sectional line perpendicular to the corresponding NHDPlus flowline was generated and mapped onto the computational raster domain. A cross-sectional width of 60~m was adopted for most gauges, while the width was extended to 90~m for four wider-channel locations. The simulated discharge values were computed by integrating the flow velocities and water depths across the intersected raster cells along each cross-section. In addition, the water surface elevation (WSE) was computed using the simulated water depth and local bed elevation and compared against the observed USGS stage measurements.

The simulated and observed hydrodynamic quantities agreed acceptably across the gauge locations. In general, the simulated water surface elevations agreed well with the observed stage measurements throughout the flood event. Larger discrepancies were observed in the discharge comparisons, particularly for gauges USGS 05528000 and USGS 05529000, where the model tended to underestimate peak flow rates. These discrepancies are attributed primarily to uncertainties associated with cross-sectional discharge integration, local hydraulic controls, and channel representation limitations at the 30~m spatial resolution. Nevertheless, the overall agreement of the water surface elevations indicates that the model adequately captures the large-scale flood propagation and inundation dynamics within the watershed.

In addition to the quantitative gauge-based validation, a qualitative flood inundation assessment was conducted using photographs collected from community social media posts during the flood event. The simulated inundation extents were visualized in three dimensions using Google Earth and compared against the observed flooded locations, as illustrated in the left panels of Figure~\ref{fig:validation}. A water-depth threshold of 15~cm was used to define inundated regions in the numerical simulations. The simulated inundation polygons agreed well with the flooded areas visible in the community photographs, in both the general inundation boundaries and the spatial flood extent, at multiple locations within the watershed.

\subsection{Historical events and training dataset}

\label{sec:dataset_generation}

A large-scale flood simulation dataset was generated to train and evaluate the proposed diffusion-based flood forecasting model. The precipitation forcing was drawn from a catalog of 120 candidate historical precipitation events identified across 12 states of the Midwestern United States (Illinois, Indiana, Iowa, Kansas, Michigan, Minnesota, Missouri, Nebraska, North Dakota, Ohio, South Dakota, and Wisconsin), spanning the period from July 2003 to September 2024. Seven events could not be retrieved completely from the precipitation archive and were discarded, leaving 113 historical events that were carried through the simulation campaign.

Precipitation forcing was obtained from the NCEP/NWS Stage IV multisensor analysis at hourly temporal resolution and approximately $4 \times 4$~km$^2$ spatial resolution. Each event was extracted over a 97-hour window spanning from one day before to three days after the nominal event date, yielding a four-day storm sequence.

To construct the precipitation forcing for the Des Plaines River watershed, each historical event was extracted over a spatial domain with the same geometric extent as the Des Plaines watershed and then projected onto the watershed raster grid. This procedure exposes the model to a diverse set of storm patterns, intensities, and spatial rainfall distributions while maintaining a consistent computational domain. It should be noted that this design deliberately transposes storms observed elsewhere in the Midwest onto the Des Plaines domain; the resulting event set is therefore physically plausible but not a reconstruction of the observed flood history of this particular watershed.

For each event, a 24-hour warm-up period was prepended to the four-day historical precipitation sequence. Simulations were initialized from a dry bed, and three different uniform precipitation intensities were prescribed during the warm-up period: 0.1, 0.5, and 1.0~mm/hr. This produced three antecedent-wetness scenarios per historical event and brought the total number of hydrodynamic simulations to $113 \times 3 = 339$ per spatial resolution. Each simulation therefore covers 121 hours in total (24~h of warm-up followed by the 97~h storm window), with the model state written out at hourly intervals.

\begin{figure}[h!]
    \centering
    \includegraphics[width=1\linewidth]{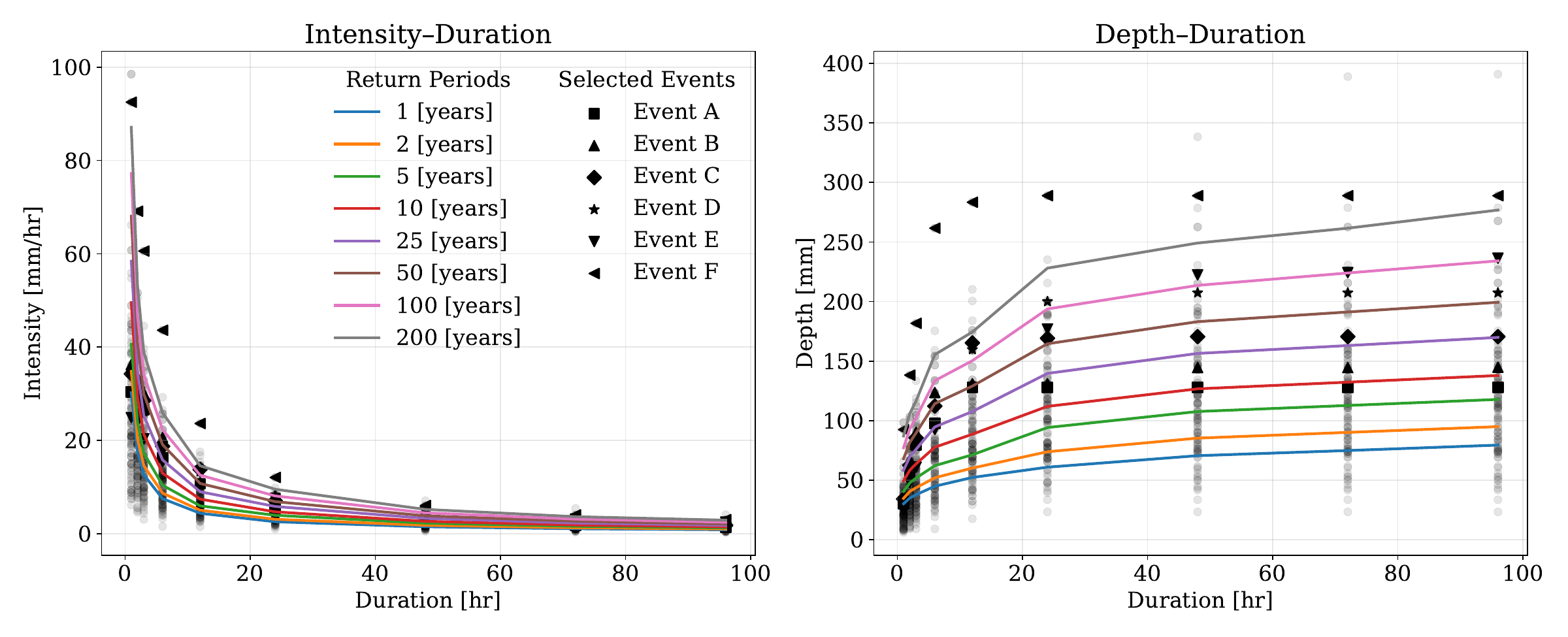}
    \caption{Intensity--Duration--Frequency (IDF) curves (left) and Depth--Duration--Frequency (DDF) curves (right) derived from NOAA Atlas 14 precipitation frequency estimates for Joliet, Illinois, within the Des Plaines River watershed. The faint scatter points represent the historical precipitation events used for the training dataset, while the solid scatter points correspond to the testing precipitation events used for model evaluation.}
    \label{fig:idf_ddf_noaa_atlas}
\end{figure}

Figure~\ref{fig:idf_ddf_noaa_atlas} presents the Intensity--Duration--Frequency (IDF) and Depth--Duration--Frequency (DDF) curves obtained from the NOAA Atlas 14 precipitation frequency estimates for Joliet, Illinois \cite{bonnin2006precipitation}. The curves cover rainfall durations ranging from 5~minutes to 4~days and return periods between 1 and 200 years. These frequency curves provide a reference framework for characterizing the severity and rarity of the precipitation events included in the generated flood simulation dataset.

The faint scatter points in Figure~\ref{fig:idf_ddf_noaa_atlas} represent the historical precipitation events used for training the diffusion model. The precipitation statistics were computed from rainfall samples extracted at the same spatial location within the projected precipitation raster over the Des Plaines River watershed. The distribution of the training events indicates that the dataset spans a broad range of precipitation severities and return periods, with a higher concentration of events corresponding approximately to the 2-year to 25-year return-period range.

The solid scatter points correspond to the six withheld testing events, denoted Events~A--F. These were selected to span the range of storm severities represented in the catalog, from Event~A, the weakest event in terms of basin-averaged accumulation, to Event~F, which exhibits by far the largest peak hourly intensity. Table~\ref{tab:test_events} summarizes the characteristics of the six testing events. It should be emphasized that storm severity is not a single scalar: the ordering of Events~A--F is monotonic in neither basin-averaged accumulation nor peak hourly intensity. Event~D produces the largest basin-averaged and basin-maximum accumulation, whereas Event~F is the most extreme in terms of short-duration intensity. The labels A--F should therefore be read as six distinct severity regimes rather than as a strictly increasing severity axis, and the metric-versus-event figures in Section~\ref{sec:results} are interpreted accordingly.

The IDF and DDF statistics for both the training and testing datasets were computed using a shifting-window analysis applied to the precipitation time series. The window duration was varied according to the target duration of interest, including short-duration and long-duration precipitation accumulations such as 5~minutes, 1~hour, and 1~day. For each duration, the maximum accumulated precipitation or rainfall intensity obtained from the shifting window analysis was retained and used for the frequency characterization.

The dataset generation was performed by running SynxFlow for all 339 event--warm-up combinations at three spatial resolutions: 30~m, 100~m, and 150~m. The 30~m simulations were used as the high-resolution target dataset, while the 100~m and 150~m simulations were used as low-resolution conditioning inputs for the diffusion model, as described in Section~\ref{sec:methodology}. All simulations were executed on eight NVIDIA A100 GPUs of the SWING cluster at the Laboratory Computing Resource Center (LCRC) of Argonne National Laboratory. The complete campaign required approximately 50.7, 7.2, and 4.1~GPU-node-hours of wall-clock time at 30~m, 100~m, and 150~m, respectively.

Patch extraction retained 111 of the 113 events for the 150~m conditioning pairing and all 113 events for the 100~m pairing, owing to two incomplete coarse-resolution outputs. The six testing events are present in both pairings, so all reported model comparisons are evaluated on an identical testing set; the training sets differ by two events (107 versus 105 training events for the 100~m and 150~m configurations, respectively).

\begin{table}[!htb]
\centering
\caption{Characteristics of the six withheld testing precipitation events, computed from the Stage~IV fields after projection onto the Des Plaines River watershed domain. Accumulations are over the 97-hour storm window.}
\label{tab:test_events}
\begin{tabular}{llllccc}
\toprule
\textbf{Event} & \textbf{Index} & \textbf{Date} & \textbf{Origin} & \textbf{Basin-mean} & \textbf{Basin-max} & \textbf{Peak 1-h} \\
 & & & & \textbf{accum. [mm]} & \textbf{accum. [mm]} & \textbf{intensity [mm/hr]} \\
\midrule
A & 40  & 2021-06-08 & North Dakota & 22.2  & 138.5 & 67.8 \\
B & 85  & 2016-08-10 & Minnesota    & 69.0  & 181.5 & 56.8 \\
C & 95  & 2015-05-06 & Nebraska     & 99.3  & 227.7 & 52.0 \\
D & 15  & 2021-06-24 & Kansas       & 131.7 & 354.5 & 63.9 \\
E & 5   & 2017-04-28 & Indiana      & 120.4 & 275.0 & 59.6 \\
F & 115 & 2018-08-20 & Wisconsin    & 80.6  & 322.0 & 145.4 \\
\bottomrule
\end{tabular}
\end{table}

\clearpage
\newpage
\section{Methodology}
\label{sec:methodology}
\subsection{Conditional denoising diffusion probabilistic model}

We employ a Denoising Diffusion Probabilistic Model (DDPM) to forecast high-resolution (HR) flood inundation depth, conditioned on static topography, observed precipitation, the initial hydrodynamic state, and optionally low-resolution (LR) inundation fields. DDPMs provide a generative framework that learns to synthesize high-resolution fields by reversing a stochastic diffusion process \cite{ho2020denoising}. As illustrated in Figure \ref{fig:forward_reverse}, diffusion models leverage a dual-stage mechanism: (i) a \textit{forward process} that systematically degrades data into Gaussian noise, and (ii) a \textit{trainable reverse process} where the model learns to reconstruct the data by iteratively removing the added noise.

The forward diffusion process $q(x_{1:T} \mid x_0)$ is a non-trainable Markov chain that progressively perturbs an HR sample $x_0$ by injecting Gaussian noise over $T$ time steps. A noisy state $x_t$ at any arbitrary noise time step $t$ can be sampled directly from $x_0$ using:
\begin{equation}
    x_t = \sqrt{\bar{\alpha}_t}x_{0} + \sqrt{1-\bar{\alpha}_t}\epsilon,
\label{eq:x_t_forward}
\end{equation}
where $\epsilon \sim \mathcal{N}(0, \mathbf{I})$ and $\bar{\alpha}_t$ is defined by the noise schedule $\beta_t$ such that $\bar{\alpha}_t = \prod_{s=1}^t (1-\beta_s)$. For a suitably chosen noise schedule, $\bar{\alpha}_T \approx 0$, so that $x_T$ is approximately distributed as a standard normal $\mathcal{N}(0, \mathbf{I})$. A sigmoid noise schedule is adopted, which concentrates the schedule's resolution at intermediate noise levels.

The reverse process $p_\theta(x_{t-1} \mid x_t, \zeta)$, conditioned on $\zeta$, approximates the intractable reverse transition $q(x_{t-1} \mid x_t)$. In this study, a convolutional neural network is trained with the $v$-prediction parameterization \cite{salimans_2022}, $v_\theta(x_t, t, \zeta) \approx v_t$, where the target $v_t$ is defined as:
\begin{equation}
    v_t \equiv \sqrt{\bar{\alpha}_t}\epsilon - \sqrt{1-\bar{\alpha}_t}x_0.
\end{equation}
This target, sometimes called the diffusion ``velocity'' and unrelated to the flow velocity $v$ of Section~\ref{sec:dataset}, balances the signal and noise components and improves training stability and sample quality, particularly for high-resolution synthesis. Given the predicted $v_\theta$, the clean sample is estimated as $\hat{x}_0 = \sqrt{\bar{\alpha}_t}x_t - \sqrt{1-\bar{\alpha}_t}v_\theta$. The reverse transition is then sampled as:
\begin{equation}
    p_\theta(x_{t-1} \mid x_t, \zeta) = \mathcal{N}\big(x_{t-1}; \mu_\theta(x_t, t, \zeta), \sigma_t^2 \mathbf{I}\big),
\end{equation}
where the mean $\mu_\theta(x_t, t, \zeta)$ is the analytical posterior mean $\tilde{\mu}(x_t, \hat{x}_0)$ evaluated using the estimated $\hat{x}_0$, and the variance $\sigma_t^2$ is set to the posterior variance $\tilde{\beta}_t$. For the Gaussian forward process, both quantities are available in closed form \cite{ho2020denoising}:
\begin{equation}
    \tilde{\mu}(x_t, \hat{x}_0) = \frac{\sqrt{\bar{\alpha}_{t-1}}\,\beta_t}{1-\bar{\alpha}_t}\,\hat{x}_0 + \frac{\sqrt{\alpha_t}\,(1-\bar{\alpha}_{t-1})}{1-\bar{\alpha}_t}\,x_t,
    \qquad
    \tilde{\beta}_t = \frac{1-\bar{\alpha}_{t-1}}{1-\bar{\alpha}_t}\,\beta_t,
\end{equation}
where $\alpha_t = 1-\beta_t$.

Sampling is performed with the full stochastic ancestral sampler over all $T$ reverse steps: the chain is initialized from $x_T \sim \mathcal{N}(0,\mathbf{I})$ and an independent Gaussian perturbation of magnitude $\sigma_t$ is injected at every step $t > 0$. No deterministic or distilled sampler is used. Consequently, each evaluation of the trained model for a fixed conditioning $\zeta$ returns an independent draw from the learned predictive distribution $p_\theta(x_0 \mid \zeta)$ rather than a single deterministic forecast. This property is exploited directly in Section~\ref{sec:ensemble_method} to construct forecast ensembles.

To guide the generation toward physically consistent flood maps, the process is conditioned on $\zeta = \{\zeta_{\text{static}}, \zeta_{\text{dynamic}}\}$, representing the environmental and hydrodynamic constraints. Detailed specifications of these conditioning variables are provided in Sections~\ref{sec:static_conditions} and \ref{sec:dynamic_conditions}.

\subsection{Forecast target and patch-based formulation}
\label{sec:target}

The quantity forecast by the model is the high-resolution water-depth field $h$ at four daily lead times. To distinguish physical time from the diffusion step $t$, lead time is denoted by $\tau$. Denoting by $\tau=0$ the end of the 24-hour warm-up period (i.e.\ the onset of the storm), the model predicts $h$ at $\tau = 24, 48, 72$ and $96$~h, which correspond to simulation hours 48, 72, 96 and 120 of the SynxFlow runs. These are instantaneous hourly model states rather than daily averages or daily maxima. The four lead times are represented as four output channels of a single network evaluation, so the entire four-day forecast is produced in one pass rather than by autoregressive roll-out. Water depth is the only forecast variable; the momentum components enter the model only as conditioning information.

Training and inference operate on spatial patches rather than on the full watershed. Candidate patches of $512 \times 512$ cells ($15.36 \times 15.36$~km$^2$ at 30~m resolution, i.e.\ 235.9~km$^2$ per patch) were extracted on a regular grid with a stride of 256 cells, corresponding to 50\% overlap in both directions. Of the 95 candidate positions, the 60 patches containing less than 50\% no-data cells in the DEM were retained. The geometric union of the retained patches covers 97.5\% of the watershed area. These 60 overlapping patches are used for training. For watershed-scale inference, the domain is tiled with 26 non-overlapping patches of the same size, so a complete forecast requires 26 patch evaluations.

All conditioning and target variables are normalized to the unit interval by min--max scaling using global statistics computed once over the entire training dataset, so that the normalization is identical for every patch and every event. Water depth is de-normalized with the same global statistics before any metric is computed or any map is plotted.

\subsection{Network architecture}
\label{sec:architecture}
\begin{figure*}[t]
    \centering
    \captionsetup[subfigure]{font=scriptsize}

    \newcommand{\figw}{0.99\linewidth}

    \begin{subfigure}{\figw}
        \centering
        \includegraphics[width=\linewidth]
        {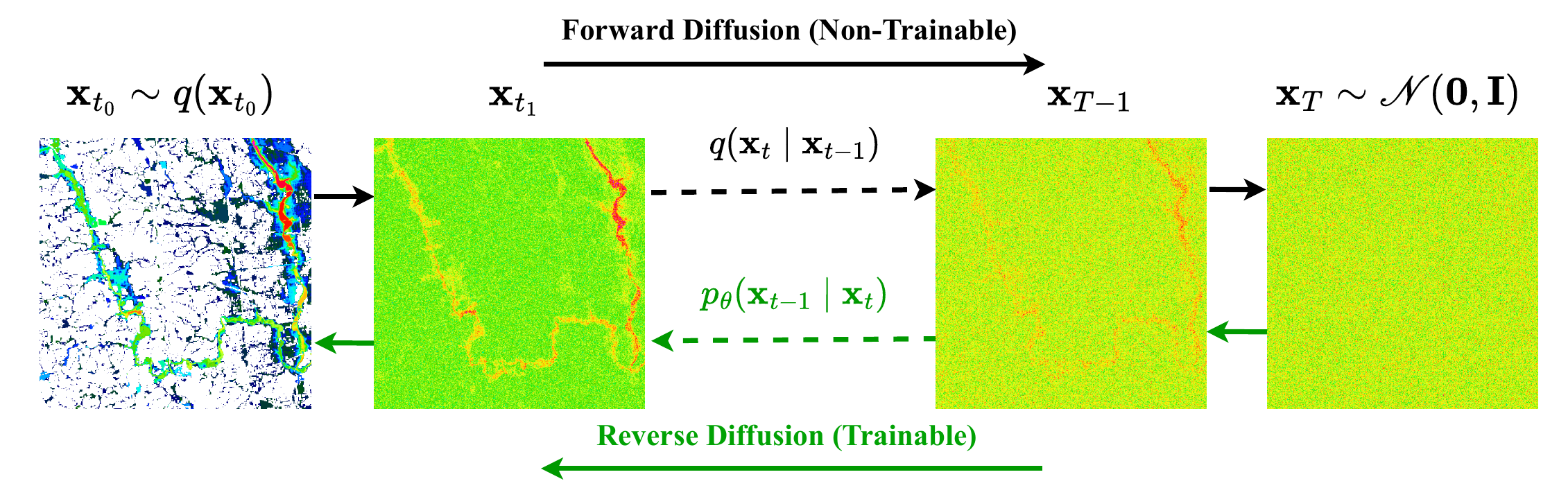}
        \caption{Forward diffusion and reverse denoising processes of the denoising diffusion probabilistic model (DDPM). The forward diffusion process is formulated as a Markovian chain that progressively adds Gaussian noise to the input data until it approaches a standard Gaussian distribution. The reverse denoising process is a trainable generative process that iteratively reconstructs the clean data sample from the noisy latent distribution.}
        \label{fig:forward_reverse}
    \end{subfigure}

    \vspace{0.4em}

    \begin{subfigure}{\figw}
        \centering
        \includegraphics[width=0.99\linewidth]
        {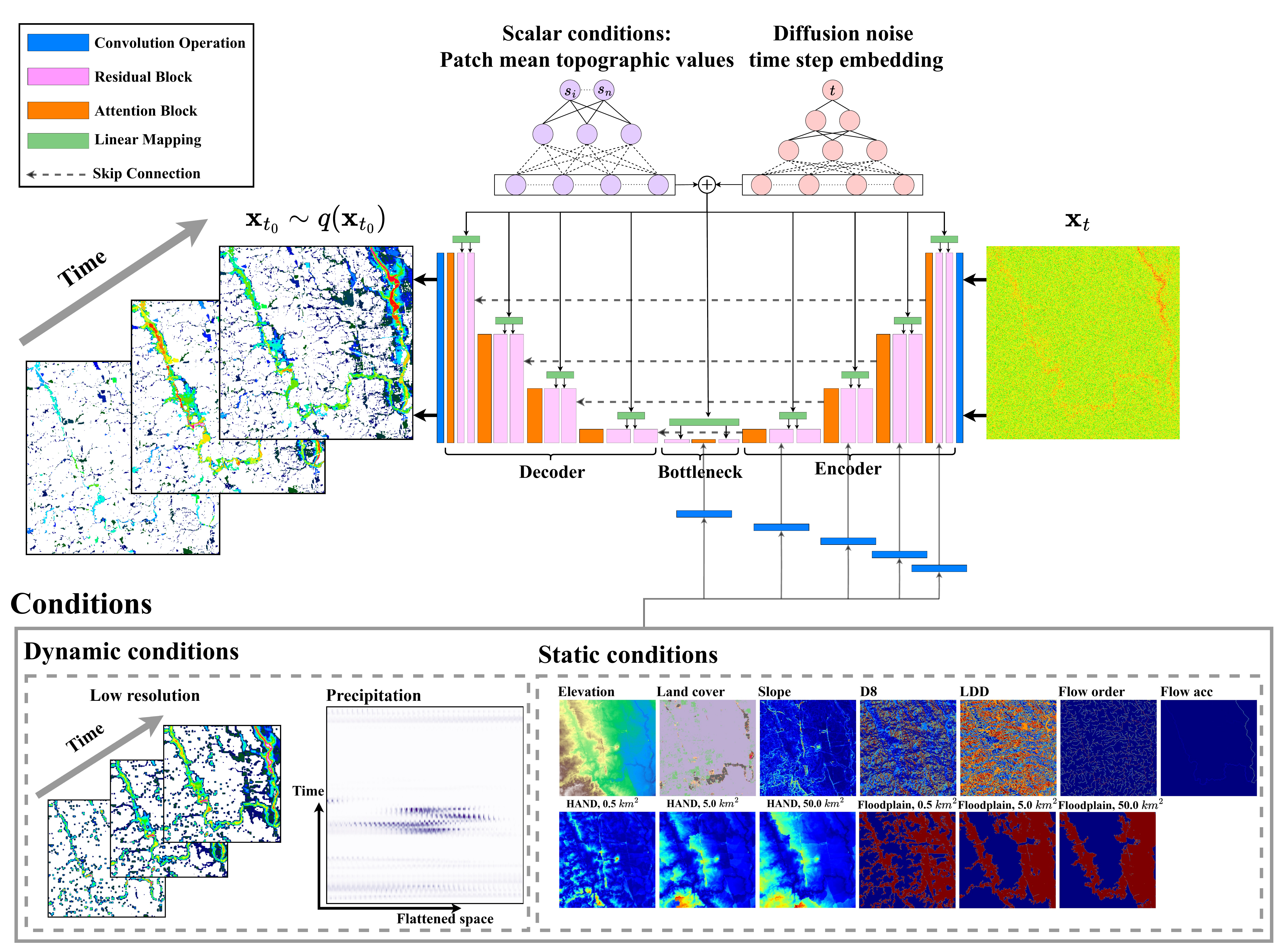}
        \caption{U-Net architecture of the denoising diffusion probabilistic model (DDPM) used for spatiotemporal flood inundation forecasting. The network predicts high-resolution spatiotemporal water-depth inundation maps from noisy latent input fields during the reverse denoising process. The DDPM is conditioned using both static and dynamic physical inputs. The static conditioning variables are derived from the digital elevation model (DEM), while the dynamic conditioning variables include the spatiotemporal precipitation forcing, the initial hydrodynamic state, and the optional low-resolution spatiotemporal flood inundation maps generated from the hydrodynamic simulations.}
        \label{fig:UNet}
    \end{subfigure}

    \caption{Conditional denoising diffusion probabilistic model architecture for spatiotemporal flood water-depth inundation forecasting.
    }
    \label{fig:DDPM}
\end{figure*}

The reverse diffusion process is parameterized by a conditional U-Net architecture \cite{ronneberger2015u}, integrating convolutional residual blocks \cite{he2016deep}, skip connections, downsampling/upsampling stages, and self-attention mechanisms \cite{zhang2019self}, as illustrated in Figure \ref{fig:UNet}.

At each reverse step, the network predicts $v_\theta$ from the noisy state $x_t$, from which the four-frame water-depth sequence $\hat{x}_0$ and the next state $x_{t-1}$ are obtained, guided by the static and dynamic conditioning set $\zeta = \{\zeta_{\text{static}}, \zeta_{\text{dynamic}}\}$.

The noisy state $x_t$ is concatenated channel-wise with the conditioning tensor, processed through an initial convolutional layer, and subsequently passed through a U-Net consisting of an encoder, bottleneck, and decoder with channel multipliers $(1,2,4,8)$. The encoder progressively increases the feature/channel depth by a factor of two at each downsampling stage. Conversely, the decoder gradually decreases the number of channels by a factor of two at each upsampling stage until the original spatial resolution is recovered, followed by a final projection to the four output channels. The base channel dimension and the number of diffusion steps were selected through the sensitivity analysis described in Section~\ref{sec:hyperparameter}.

The composition of the conditioning tensor is summarized in Table~\ref{tab:channels}. In addition to these two-dimensional spatial conditioning tensors, four scalar conditioning variables are incorporated into the network through a dedicated embedding network, together with the diffusion time step $t$. The diffusion time step informs the model of the current noise level within the diffusion process, allowing the denoising network to adapt its predictions accordingly. The four scalars are the patch-mean values of the normalized elevation, land cover, slope, and large-threshold HAND layers, and provide global patch-wise topographic context that complements the purely local receptive field of the convolutional encoder at the coarsest scales. We note that the patch mean of the land-cover layer is a summary of a categorical variable and should be interpreted as a coarse proxy for the degree of urbanization within the patch rather than as a physical quantity.

The four scalar patch descriptors and the diffusion step $t$ are processed by independent multi-layer perceptrons (MLPs). The resulting embeddings are then injected into every residual block of the U-Net (Figure \ref{fig:UNet}) to ensure global context is preserved across all spatial scales.

\begin{table}[!htb]
\centering
\caption{Composition of the input and output tensors of the conditional U-Net for a $512\times512$ patch. The low-resolution block is present only in the LR-conditioned configurations.}
\label{tab:channels}
\begin{tabular}{llcp{0.46\linewidth}}
\toprule
\textbf{Role} & \textbf{Block} & \textbf{Ch.} & \textbf{Description} \\
\midrule
Target  & $x_0$                   & 4  & Water depth $h$ at lead times 24, 48, 72, 96~h \\
\midrule
\multirow{4}{*}{Conditioning}
        & $\zeta_{\text{static}}$ & 13 & Terrain and land-surface layers (Figure~\ref{fig:static_conditions}) \\
        & $\zeta_{\text{init}}$   & 3  & $[h, hu, hv]$ at the end of the warm-up ($\tau=0$) \\
        & $\zeta_{\text{precip}}$ & 1  & Stage~IV spatiotemporal precipitation matrix \\
        & $\zeta_{\text{LR}}$     & 15 & 5 time levels $\times$ $[h, hu, hv]$ at 100 or 150~m \\
\midrule
        & Total (with LR)         & 32 & \\
        & Total (without LR)      & 17 & \\
\midrule
Scalars & $t$, patch means        & $1{+}4$ & Diffusion step; patch means of elevation, land cover, slope and HAND \\
\bottomrule
\end{tabular}
\end{table}

\subsection{Static conditions}
\label{sec:static_conditions}
\begin{figure*}[t]
    \centering
    \captionsetup[subfigure]{font=scriptsize}

    \newcommand{\figw}{0.13\linewidth}

    \begin{subfigure}{\figw}
        \includegraphics[width=\linewidth]{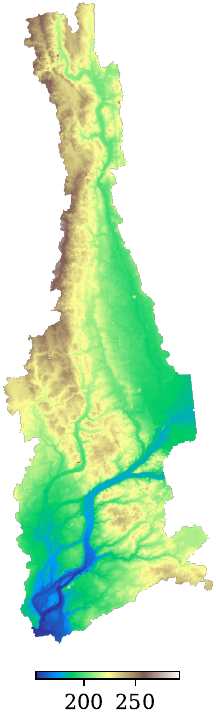}
        \caption{Elevation}
    \end{subfigure}
    \hfill
    \begin{subfigure}{\figw}
        \includegraphics[width=\linewidth]{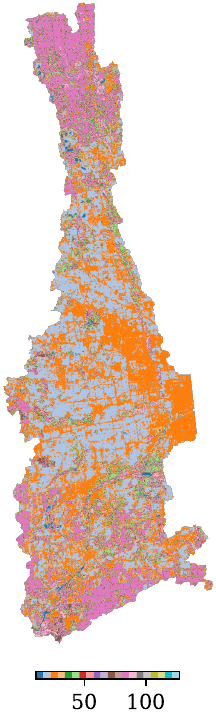}
        \caption{Land cover}
    \end{subfigure}
    \hfill
    \begin{subfigure}{\figw}
        \includegraphics[width=\linewidth]{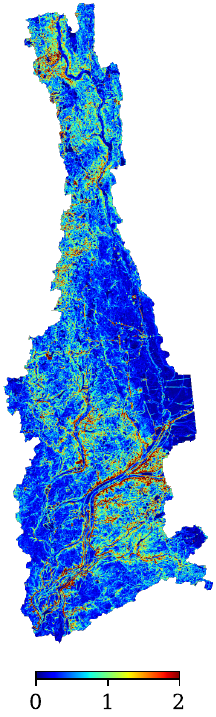}
        \caption{Slope}
    \end{subfigure}
    \hfill
    \begin{subfigure}{\figw}
        \includegraphics[width=\linewidth]{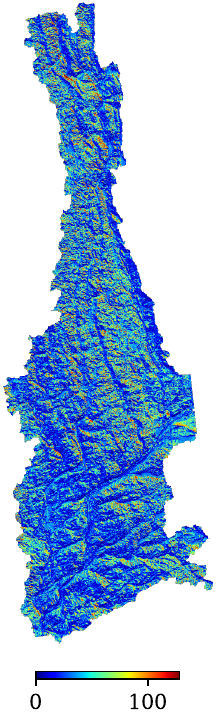}
        \caption{D8}
    \end{subfigure}
    \hfill
    \begin{subfigure}{\figw}
        \includegraphics[width=\linewidth]{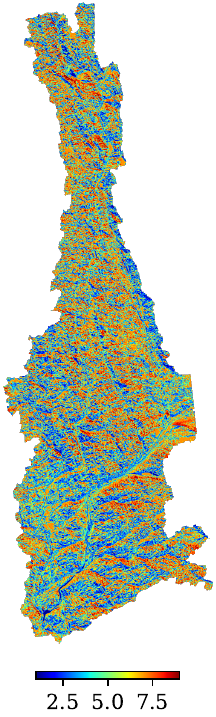}
        \caption{LDD}
    \end{subfigure}
    \hfill
    \begin{subfigure}{\figw}
        \includegraphics[width=\linewidth]{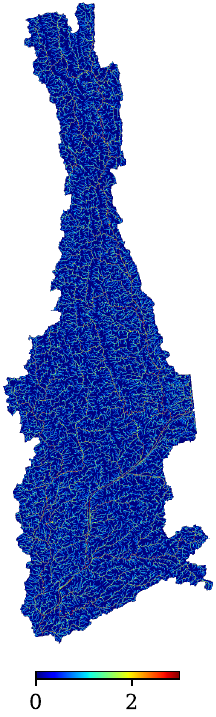}
        \caption{Stream order}
    \end{subfigure}
    \hfill
    \begin{subfigure}{\figw}
        \includegraphics[width=\linewidth]{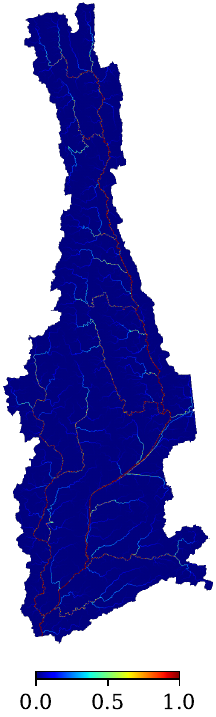}
        \caption{Acc $\times 10^6$ }
    \end{subfigure}

    \vspace{0.4em}

    \begin{subfigure}{\figw}
        \includegraphics[width=\linewidth]{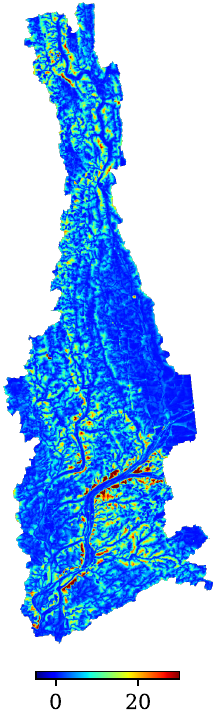}
        \caption{HAND, $0.5\ \mathrm{km}^2$}
    \end{subfigure}
    \hfill
    \begin{subfigure}{\figw}
        \includegraphics[width=\linewidth]{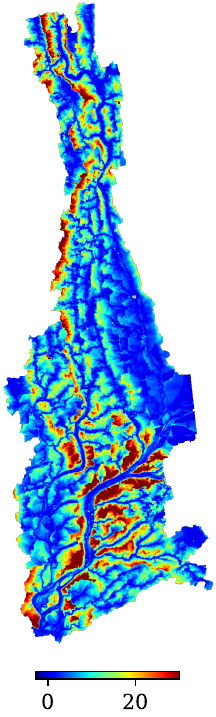}
        \caption{HAND, $5.0\ \mathrm{km}^2$}
    \end{subfigure}
    \hfill
    \begin{subfigure}{\figw}
        \includegraphics[width=\linewidth]{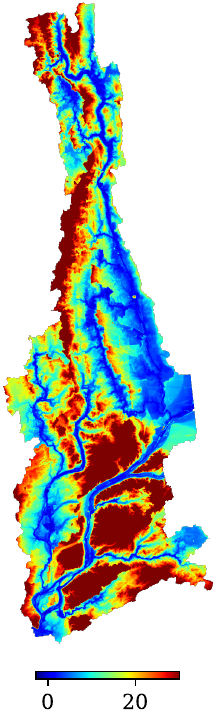}
        \caption{HAND, $50.0\ \mathrm{km}^2$}
    \end{subfigure}
    \hfill
    \begin{subfigure}{\figw}
        \includegraphics[width=\linewidth]{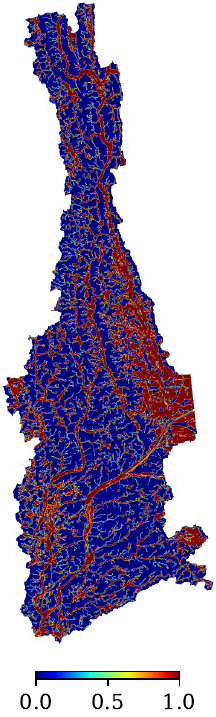}
        \caption{Floodplain, $0.5\ \mathrm{km}^2$}
    \end{subfigure}
    \hfill
    \begin{subfigure}{\figw}
        \includegraphics[width=\linewidth]{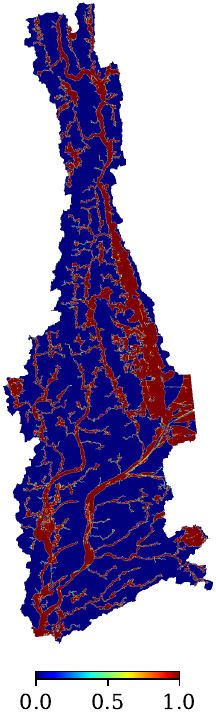}
        \caption{Floodplain, $5.0\ \mathrm{km}^2$}
    \end{subfigure}
    \hfill
    \begin{subfigure}{\figw}
        \includegraphics[width=\linewidth]{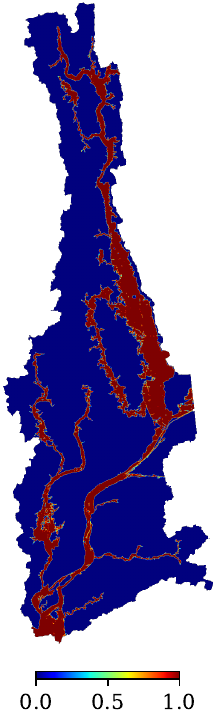}
        \caption{Floodplain, $50.0\ \mathrm{km}^2$}
    \end{subfigure}

    \caption{Static conditioning variables used by the conditional DDPM framework. The digital elevation model (DEM) was obtained from the USGS 3D Elevation Program (3DEP) at 30~m spatial resolution, and the land-cover data were obtained from the National Land Cover Database (NLCD) 2021 at 30~m resolution. The remaining geomorphic and hydrologic conditioning variables, including slope, flow direction, stream order, flow accumulation, HAND, and geomorphic floodplain layers, were derived from the DEM using the \texttt{pyflwdir} package \cite{DirkEilander2021}.}

    \label{fig:static_conditions}
\end{figure*}

The static conditioning variables, denoted by $\zeta_{\text{static}}$, are derived from the watershed topography and land-surface characteristics and are used to provide spatial information governing runoff. The adopted static conditioning layers are illustrated in Figure~\ref{fig:static_conditions} and comprise 13 channels: elevation, land cover, slope, local flow direction in the D8 and LDD encodings, stream order, flow accumulation, Height Above the Nearest Drainage (HAND), and geomorphic floodplain maps. The HAND and geomorphic floodplain raster sets were each generated using three different upstream-area thresholds of $\{0.5, 5.0, 50.0\}$ km$^2$ in order to capture drainage characteristics at multiple spatial scales, which accounts for six of the thirteen channels.

The elevation data were obtained from the USGS 3D Elevation Program (3DEP) digital elevation model at 30~m spatial resolution, while land-cover information was obtained from the National Land Cover Database (NLCD) 2021 dataset at the same spatial resolution. Prior to deriving the topographic variables, surface depressions were removed through a pit-filling procedure to ensure hydrologically consistent drainage networks. The pit-filling algorithm implemented in the \texttt{pyflwdir} package follows the efficient depression-filling method proposed in \cite{wang2006efficient}.

Several terrain-derived hydrologic variables were computed using the open-source Python package \texttt{pyflwdir} \cite{DirkEilander2021}. The slope was computed from the DEM using first-order partial derivatives estimated by finite differences within a moving $3\times3$ cell window and represents the local terrain gradient expressed as elevation change per unit horizontal distance (m/m).

The local flow direction was encoded using both the D8 and the PCRaster Local Drain Direction (LDD) conventions, which represent the same single-direction drainage network using different numeric codes. These variables describe the downstream connectivity between neighboring cells and define the direction of surface water movement across the terrain. The downstream flow path is determined using the steepest-descent method, where each cell drains toward the neighboring cell with the steepest downslope gradient. The flow-direction network was derived from the depression-filled DEM obtained with the method of \cite{wang2006efficient}. Because D8 and LDD are alternative encodings of the same drainage network, these two channels are mutually redundant; they are both retained here for consistency with the archived dataset, and their redundancy is noted as a minor inefficiency rather than an additional source of information.

Based on the derived flow-direction network, stream order and flow accumulation maps were computed. Stream order was calculated using the Strahler stream-order classification system \cite{strahler1957quantitative}, where first-order streams represent headwater channels. When two streams of the same order merge, the downstream segment increases by one order, whereas the confluence of streams with different orders inherits the larger stream order. Flow accumulation represents the total contributing upstream area draining through each cell and is widely used as a proxy for channel initiation, drainage density, and watershed connectivity.

The HAND layers were generated in \texttt{pyflwdir} following the methodology described in~\cite{nobre2016hand}, which computes the vertical distance between a terrain cell and its nearest downstream drainage channel. By normalizing topography relative to the local drainage network, HAND provides a physically meaningful indicator of flood susceptibility and potential inundation extent. Cells with lower HAND values generally exhibit higher flood vulnerability because they lie at lower elevations relative to the local drainage network.

In addition, geomorphic floodplain maps were generated using the GFPLAIN approach proposed in~\cite{nardi2019gfplain250m}. This method delineates floodplain boundaries using a power-law (hydraulic scaling) relationship between upstream contributing area and flow depth, which defines the maximum HAND value considered part of the floodplain. The resulting floodplain layers provide information regarding areas that are geomorphologically prone to flooding and have been shown to be effective predictors of flood extent in data-driven hydrologic applications.

Together, these static conditioning variables provide complementary information describing the watershed terrain, drainage structure, and flood susceptibility, thereby enabling the DDPM framework to learn the spatial controls governing flood inundation dynamics.

\subsection{Dynamic conditions}
\label{sec:dynamic_conditions}

The dynamic conditioning set $\zeta_{\text{dynamic}} = \{\zeta_{\text{init}}, \zeta_{\text{precip}}, \zeta_{\text{LR}}\}$ comprises three blocks:

\begin{itemize}
    \item \textbf{Initial hydrodynamic state ($\zeta_{\text{init}}$):} The water depth and the two unit-width discharge components, $[h, hu, hv]^{T}$, taken from the 30~m simulation at the end of the 24-hour warm-up period, i.e.\ at the instant the storm begins. This block defines the antecedent wetness state of the patch and is the mechanism through which the three warm-up intensity scenarios enter the model.

    \item \textbf{Precipitation ($\zeta_{\text{precip}}$):} A single-channel representation of the 97-hour hourly Stage~IV precipitation sequence over the watershed. The hourly precipitation block, of shape (97 hours $\times$ $n_y$ $\times$ $n_x$) in units of mm/hr, is flattened over its two spatial dimensions and resampled onto a $512 \times 512$ matrix whose first axis corresponds to time and whose second axis corresponds to the flattened spatial index, as sketched in Figure~\ref{fig:UNet}. This encoding supplies the network with the full storm hyetograph and its spatial heterogeneity in a single channel of the same shape as the other conditioning layers. Two consequences should be stated explicitly. First, because the Stage~IV grid is approximately 4~km while the target grid is 30~m, the precipitation channel carries the storm structure at its native mesoscale resolution only. Second, the channel is constructed once per event and is therefore identical for every patch: precipitation enters the model as a domain-wide spatiotemporal descriptor rather than as a patch-local rainfall field. Patch-local forcing information is instead conveyed through $\zeta_{\text{init}}$ and, when available, $\zeta_{\text{LR}}$.

    \item \textbf{Low-resolution hydrodynamic solution ($\zeta_{\text{LR}}$):} The coarse-grid hydrodynamic solution generated by SynxFlow at 100~m or 150~m resolution, comprising five time levels (the warm-up end state plus the four forecast lead times) of the three hydrodynamic variables $[h, hu, hv]^{T}$, for 15 channels in total. For each patch the coarse field is cropped to the corresponding geographic window and bilinearly upsampled onto the $512 \times 512$ target grid. The LR frames are time-coincident with the initial state and the four target frames, so no temporal interpolation or offset is involved. Notably, $\zeta_{\text{LR}}$ is an optional modality: the architecture retains forecasting capability using only static, initial-state, and precipitation conditioning when low-resolution simulations are unavailable, which is the configuration referred to throughout as ``No LR''.
\end{itemize}

\subsection{Ensemble construction}
\label{sec:ensemble_method}

Deterministic machine-learning flood surrogates return a single field and convey no information about forecast confidence, which is precisely the information an emergency manager needs when deciding whether to evacuate a neighborhood. Ensemble prediction is the established remedy in both meteorology \cite{leutbecher2008ensemble} and operational flood forecasting \cite{cloke2009ensemble}.

The diffusion formulation adopted here provides this capability without any additional machinery. Because the reverse process is the stochastic Markov chain described above, the trained network does not define a function from conditioning to forecast but a conditional distribution $p_\theta(x_0 \mid \zeta)$ from which independent samples can be drawn. Evaluating the identical trained model $N$ times with the identical conditioning tensor $\zeta$ yields $N$ distinct, equally plausible high-resolution flood evolutions. This is a structural property of the model class rather than an extension of it: the ensembles reported in Section~\ref{sec:results} required no retraining, no perturbation of the inputs, no auxiliary variance network, and no change to the architecture or the trained weights. It is also the principal practical advantage of diffusion models over the generative adversarial networks discussed in Section~\ref{sec:introduction}, which are prone to mode collapse and therefore tend to under-represent the diversity of plausible outcomes. Diffusion-based ensembles have recently been shown to be competitive with operational physics-based ensembles for global weather prediction \cite{price2025probabilistic}.

In practice, ensembles were generated by replicating the conditioning tensor $N = 20$ times along the batch dimension and issuing a single batched call to the ancestral sampler. Independence between members follows from the sampler drawing its initial state $x_T$ and every intermediate increment independently per batch element.

\subsubsection*{Sources of uncertainty represented by the ensemble}

It is important to delimit the source of uncertainty that is sampled, because these ensembles are not a complete predictive uncertainty budget.

Because the conditioning tensor $\zeta$ is held fixed across members, the spread represents the uncertainty of the learned generative mapping itself: given a particular terrain, antecedent state, storm and, where used, coarse hydrodynamic solution, it quantifies how much freedom the model retains in placing fine-scale water. This combines the irreducible ambiguity of reconstructing 30~m inundation detail that the conditioning does not determine with the epistemic uncertainty of the trained network. It is the appropriate uncertainty to report for the surrogate step itself, and it is the component that a deterministic surrogate silently discards.

Three further sources are deliberately \emph{not} sampled. Meteorological uncertainty is excluded, because the model is driven by observed Stage~IV precipitation rather than by a precipitation forecast. Hydrodynamic parameter uncertainty is excluded, because Manning roughness is prescribed deterministically from land cover. Antecedent-condition uncertainty is represented only coarsely, through the three discrete warm-up scenarios.

A complete operational uncertainty budget follows naturally from the same framework, and its structure is what makes the approach attractive. Given a $K$-member numerical weather prediction precipitation ensemble, each member $k$ defines its own conditioning tensor $\zeta_k$, from which $M$ diffusion members can be drawn, giving a nested $K \times M$ ensemble that propagates meteorological and generative uncertainty jointly. Its cost is dominated by the $K$ coarse hydrodynamic simulations rather than by the $KM$ diffusion samples, as quantified in Section~\ref{sec:cost}. Establishing the skill of such a system requires forecast precipitation and is left to future work.

\subsection{Interpretation of the two operating modes}
\label{sec:operating_modes}

The presence or absence of $\zeta_{\text{LR}}$ defines two conceptually distinct operating modes, and it is important to state what each one is, because they answer different questions.

In the LR-conditioned mode the coarse hydrodynamic solution is supplied at the same lead times as the quantities being predicted. The model is therefore not forecasting the flood evolution independently; it is acting as a learned \emph{spatial downscaling operator} that maps a coarse-resolution hydrodynamic forecast onto the 30~m grid, while adding the fine-scale inundation structure that the coarse solution cannot resolve. In an operational setting this corresponds to a two-stage pipeline in which a cheap coarse SWE model is driven by forecast precipitation and the diffusion model then refines its output. The skill reported for this mode should be read as the fidelity of that refinement step, conditional on the coarse forecast being available, and not as evidence that the network has learned to integrate the shallow water equations forward in time.

In the unconditioned (No-LR) mode, by contrast, the model receives only static terrain descriptors, the antecedent hydrodynamic state at the storm onset, and the precipitation sequence. No hydrodynamic information about the forecast period is provided. This mode is therefore a genuine end-to-end surrogate that maps forcing and initial state directly to the four-day inundation evolution, and it is the configuration that would be deployed if no hydrodynamic model were run at all. Its substantially larger errors, quantified in Section~\ref{sec:results}, measure the difficulty of that harder task rather than a failure of the architecture.

Both modes are reported throughout, and the contrast between them is the quantitative basis for the physics-guided conditioning argument advanced in this paper.

\subsection{Hyperparameter selection and training}
\label{sec:hyperparameter}
For the DDPM architecture shown in Figure~\ref{fig:DDPM}, a hyperparameter sensitivity analysis was conducted to select the number of diffusion denoising time steps and the base number of channels in the U-Net architecture. Six configurations were trained, covering all combinations of $\{100, 250, 500\}$ diffusion steps and $\{32, 64\}$ base channels.

To compare the configurations on an equal footing, all models were trained for a fixed wall-clock time of four hours with the same GPU hardware, learning rate, and batch size.

\begin{figure}
    \centering
    \includegraphics[width=1\linewidth]{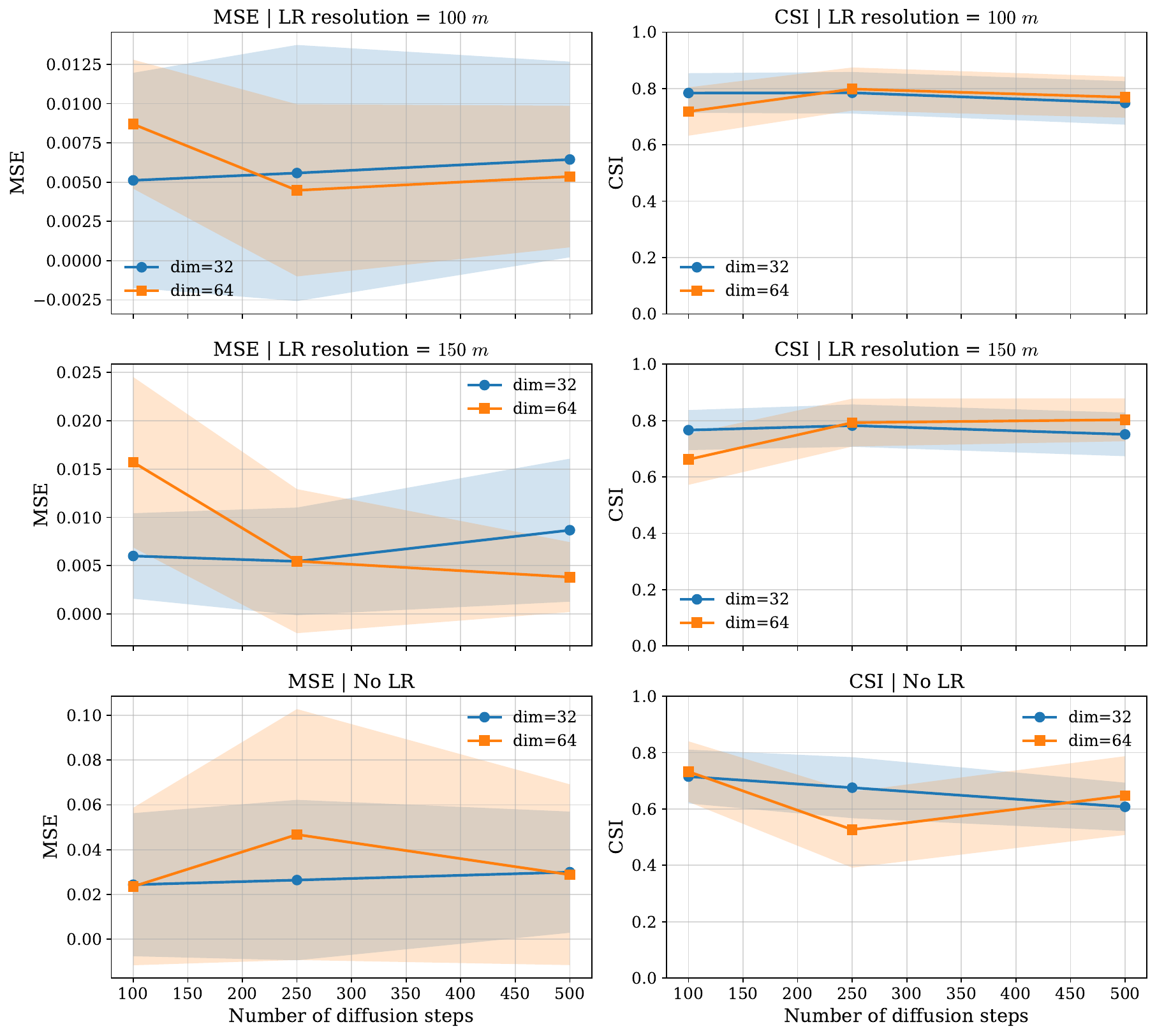}
    \caption{Water-depth mean squared error (MSE) (left) and critical success index (CSI) (right) for different combinations of diffusion denoising time steps and U-Net base channel dimensions used in the DDPM architecture shown in Figure \ref{fig:DDPM}.}
    \label{fig:tuning}
\end{figure}

The configurations were compared using the water-depth mean squared error (MSE) and the critical success index (CSI), for the 100~m, 150~m and unconditioned configurations (Figure~\ref{fig:tuning}).

The results show that for the larger U-Net base channel dimension of 64, increasing the number of diffusion denoising time steps from 100 to 500 generally improves the forecasting performance, resulting in lower MSE values and higher CSI values. In contrast, for the smaller base channel dimension of 32, increasing the number of diffusion denoising time steps tends to increase the MSE and reduce the CSI values. Under the equal-wall-clock training protocol adopted here, larger configurations complete fewer optimizer updates in the allotted four hours, and this confounding should be kept in mind when interpreting Figure~\ref{fig:tuning}: the comparison reflects accuracy attainable per unit of training budget rather than accuracy at convergence.

In particular, the configuration with 32 base channels and 100 diffusion steps performs comparably to that with 64 base channels and 500 steps. Since increasing either the number of diffusion denoising time steps or the number of U-Net channels significantly increases the inference time of a four-day watershed-scale forecast on the same GPU hardware, as summarized in Table~\ref{tab:time_tuning}, the final DDPM configuration adopted in this study utilized 100 diffusion denoising time steps for both training and inference and a U-Net base channel dimension of 32.

For the flood prediction results presented in Section \ref{sec:results}, three models were trained, one per conditioning configuration (100~m LR, 150~m LR, and No LR). Each model was trained with the Adam optimizer \cite{kingma_adam_2014} at a constant learning rate of $5 \times 10^{-5}$, a batch size of 4 patches, and mixed-precision (FP16) arithmetic, using 100 diffusion denoising time steps and a U-Net base channel dimension of 32. Training ran for 49 epochs for the 100~m configuration and 50 epochs for the 150~m and No-LR configurations, by which point the training loss had plateaued near $1\times10^{-4}$. The reported model for each configuration is the checkpoint attaining the lowest epoch-averaged training loss. Numerical experiments were conducted on the SWING GPU cluster at the Laboratory Computing Resource Center (LCRC) at Argonne National Laboratory, utilizing a single NVIDIA A100 GPU per training run.

\begin{table}[!htb]
\centering
\caption{DDPM inference time for a complete four-day watershed-scale flood forecast (26 non-overlapping patches of $512\times512$ cells) on a single NVIDIA A100 GPU, obtained from the measured per-patch inference time, for different combinations of U-Net base channel dimensions and diffusion denoising time steps. Values are averaged over the three conditioning configurations, whose timings differ by less than 2\%.}
\label{tab:time_tuning}
\begin{tabular}{ccc}
\toprule
\textbf{Base Channels} & \textbf{Diffusion Steps} & \textbf{Inference Time [sec]} \\
\midrule
32 & 100 & 123 \\
32 & 250 & 292 \\
32 & 500 & 578 \\
\midrule
64 & 100 & 163 \\
64 & 250 & 390 \\
64 & 500 & 743 \\
\bottomrule
\end{tabular}
\end{table}

\clearpage
\newpage
\section{Results and Discussion}
\label{sec:results}

All forecasts reported in this section are \emph{ensemble} forecasts. As described in Section~\ref{sec:ensemble_method}, the reverse diffusion process is stochastic, so a single evaluation of the trained model returns one arbitrary draw from the learned predictive distribution and is not by itself a well-defined forecast product. We therefore report the 20-member ensemble throughout: the ensemble mean where a single field is required, and the ensemble distribution where uncertainty is the quantity of interest. Single-realization results are not presented as forecast products.

Three conditioning configurations are compared: the DDPM conditioned on 100~m and on 150~m low-resolution SynxFlow simulations, and the unconditioned configuration (``No LR'') that receives only static terrain, the antecedent hydrodynamic state, and precipitation. All results are evaluated against the 30~m SynxFlow simulations, which serve as the reference.

\begin{figure}[h!]
    \centering
    \includegraphics[width=1\linewidth]{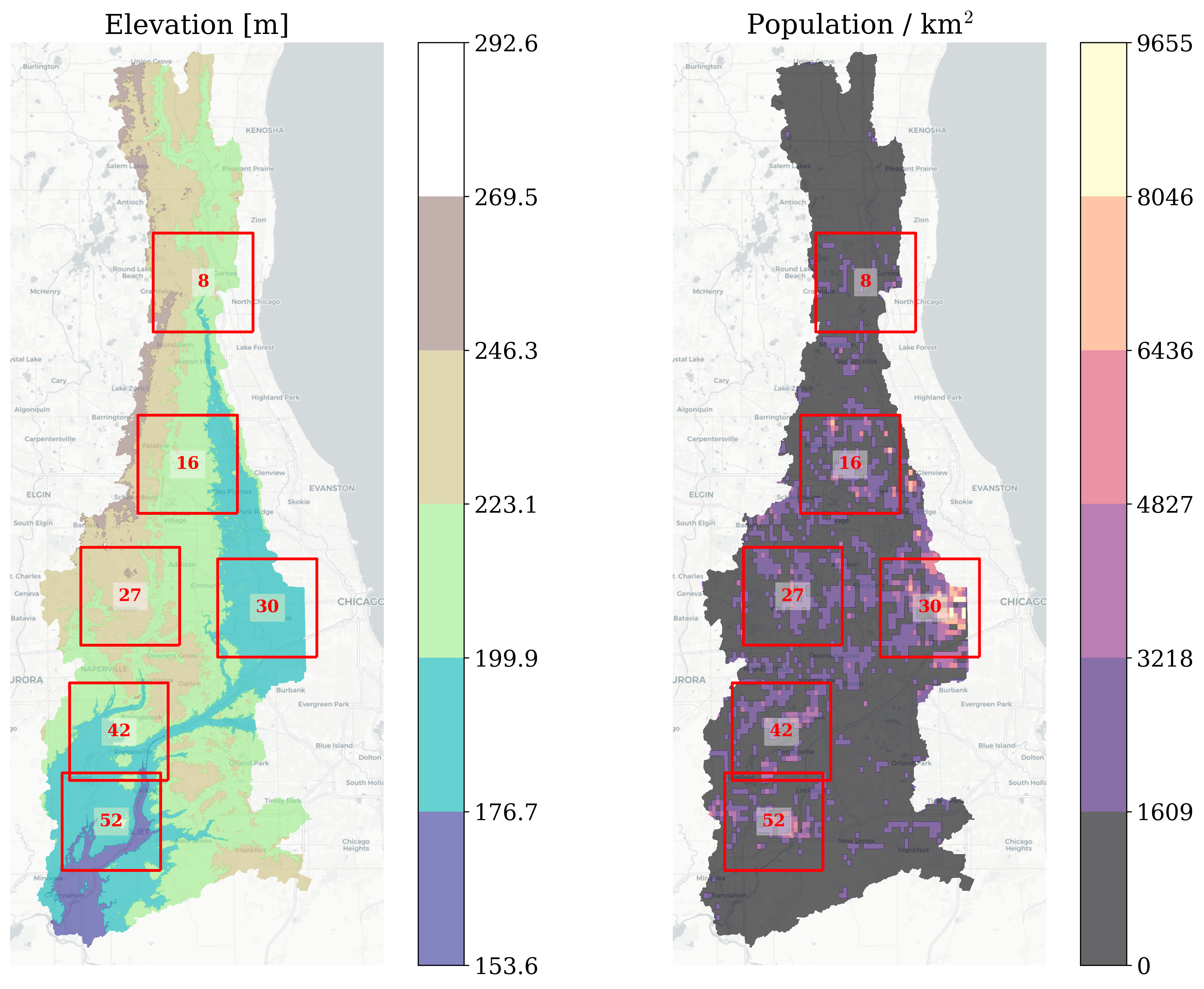}
    \caption{The six analysis patches used throughout this section, shown on the watershed terrain elevation (left) and on population density per $\mathrm{km}^2$ (right). The patches were selected for high population density and urban exposure and span the northern, central and southern reaches of the Des Plaines River watershed.}
    \label{fig:patch_dem_pop}
\end{figure}

\paragraph{Scope of the evaluation.} The ensemble experiment covers the six withheld testing events of Table~\ref{tab:test_events}, the six analysis patches highlighted in Figure~\ref{fig:patch_dem_pop}, and the 0.1~mm/hr antecedent-wetness scenario, for $3 \times 6 \times 6 \times 20 = 2160$ four-day patch forecasts. The six patches were selected for their high population density and urban exposure and together span the northern, central and southern reaches of the watershed. They are a subset of the 60 overlapping patches used for training, so the aggregate scores below are basin-representative rather than basin-complete; Section~\ref{sec:agg} shows that they reproduce the full-watershed aggregate closely. Unless stated otherwise, a cell is classified as inundated when its water depth exceeds 0.15~m, applied identically to forecasts and to the reference.

\subsection{Spatiotemporal evolution of the forecast inundation}

\begin{figure}[h!]
    \centering
    \includegraphics[width=0.99\linewidth]{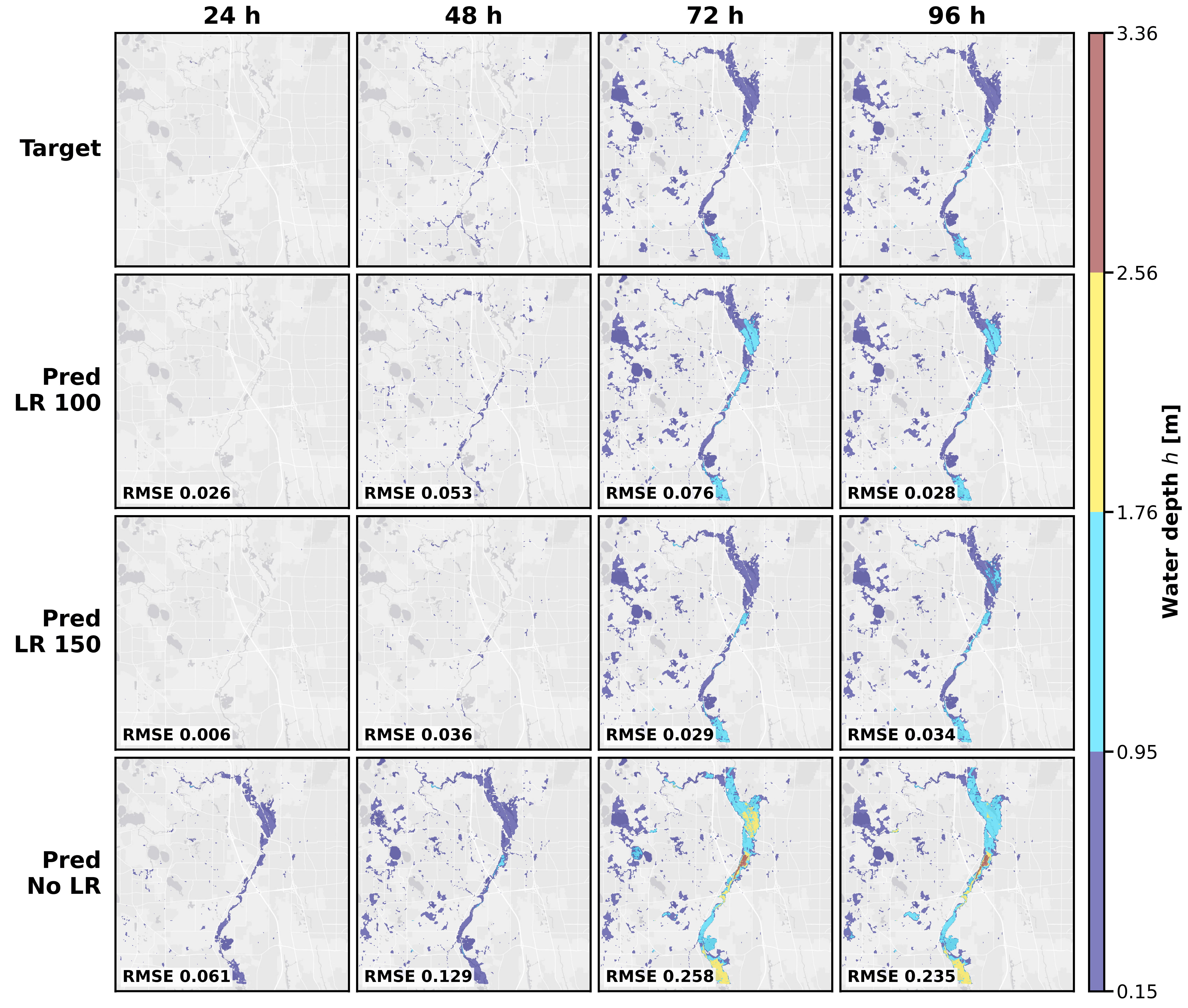}
    \caption{Ensemble-mean water-depth evolution for Event~F at patch~8. Rows: 30~m SynxFlow reference, then the ensemble mean ($N=20$) of the 100~m conditioned, 150~m conditioned and unconditioned DDPM. Columns: lead times of 24, 48, 72 and 96~h. Depths below the 0.15~m inundation threshold are transparent; the annotated RMSE is that of the ensemble mean against the reference.}
    \label{fig:days_8}
\end{figure}

Figures~\ref{fig:days_8} through \ref{fig:days_52} present the ensemble-mean water-depth fields for the six analysis patches under Event~F, the testing event with the largest peak hourly rainfall intensity (145.4~mm/hr, Table~\ref{tab:test_events}). Each figure compares the 30~m reference against the ensemble mean of the three conditioning configurations at lead times of 24, 48, 72 and 96~h.

The conditioned ensembles track the reference closely in both inundation extent and timing. Channel inundation along the main stem appears at the correct lead time, the floodplain fills in the correct sequence, and the recession between 72 and 96~h is reproduced. Per-panel RMSE values annotated on the figures remain below roughly 0.09~m at every lead time for both conditioned configurations.

The unconditioned ensemble behaves qualitatively differently, and in a consistent direction: it floods too early and too widely. For Patch~8 (Figure~\ref{fig:days_8}) it already shows extensive channel and floodplain inundation at 24~h, when the reference is almost entirely dry, and by 72~h its RMSE reaches 0.26~m against 0.076~m and 0.029~m for the 100~m and 150~m conditioned ensembles. The same premature, over-extensive pattern recurs across the other patches. Because the unconditioned model receives no hydrodynamic information about the forecast period, it must infer the timing of the flood wave from the precipitation descriptor and the antecedent state alone, and it evidently resolves that ambiguity by wetting the geomorphic floodplain early rather than by propagating a wave.

This is the central qualitative result: low-resolution hydrodynamic conditioning supplies the temporal structure that the generative model cannot recover from forcing alone, while the diffusion model supplies the fine-scale spatial detail that the coarse simulation cannot resolve.

\foreach \p in {16,27,30,42,52} {
    \begin{figure}[h!]
        \centering
        \includegraphics[width=0.99\linewidth]{figures/DM/ensemble/days_patch_\p.png}
        \caption{Ensemble-mean water-depth evolution for Event~F at patch~\p, plotted as in Figure~\ref{fig:days_8}.}
        \label{fig:days_\p}
    \end{figure}
}

\subsection{Flood-extent skill}

\begin{figure}[h!]
    \centering
    \includegraphics[width=0.99\linewidth]{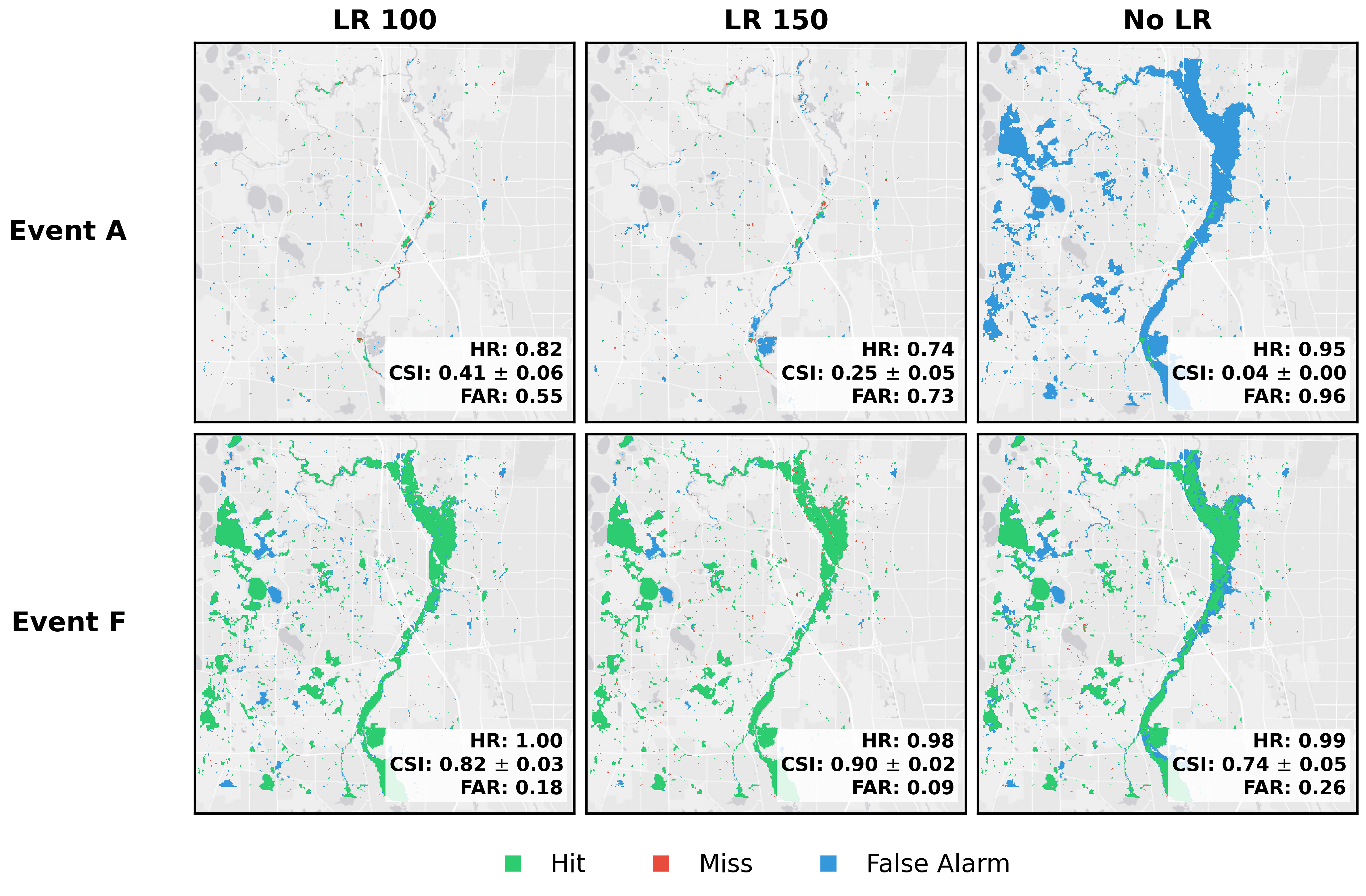}
    \caption{Ensemble-mean flood extent decomposed into hits, misses and false alarms at patch~8, on the day of maximum reference inundation. Rows: Events~A and F. Columns: the three conditioning configurations. Annotations give the hit rate, critical success index (with the standard deviation of CSI across the 20 members) and false alarm ratio of the ensemble mean.}
    \label{fig:cat_8}
\end{figure}

Figures~\ref{fig:cat_8} through \ref{fig:cat_52} resolve the forecast flood extent into hits, misses and false alarms for the weakest event by basin-averaged accumulation (Event~A) and the event with the largest peak intensity (Event~F), at the day of maximum reference inundation. The contingency metrics are defined as
\begin{equation}
    HR = \frac{TP}{TP + FN}, \qquad
    MR = \frac{FN}{TP + FN} = 1 - HR, \qquad
    FAR = \frac{FP}{TP + FP}, \qquad
    CSI = \frac{TP}{TP + FP + FN},
\end{equation}
where $TP$, $FP$ and $FN$ are the counts of cells flooded in both forecast and reference, in the forecast only, and in the reference only. Counts are taken over cells where both fields are valid. Each panel is annotated with the scores of the ensemble mean and, for CSI, with the standard deviation of that score across the 20 individual members, which indicates how much the skill of any one realization varies.

Two regimes are evident. For the weak Event~A the inundated fraction of the domain is small and all three configurations over-predict extent, but to very different degrees: at patch~8 the unconditioned ensemble reaches $FAR = 0.96$ and $CSI = 0.04$, against $FAR = 0.55$, $CSI = 0.41$ for the 100~m conditioning and $FAR = 0.73$, $CSI = 0.25$ for the 150~m conditioning. Weak events are the hardest case for all configurations, which is consistent with their under-representation in the training catalog (Figure~\ref{fig:idf_ddf_noaa_atlas}).

For the severe Event~F the picture reverses and the scores become high and well separated: hit rates approach unity for every configuration, misses are rare, and the configurations are distinguished almost entirely by their false alarms. At patch~8 the 150~m conditioned ensemble attains $CSI = 0.90$ with $FAR = 0.09$, against $CSI = 0.82$, $FAR = 0.18$ at 100~m and $CSI = 0.74$, $FAR = 0.26$ unconditioned. The member-to-member standard deviation of CSI is small throughout, between 0.02 and 0.06, confirming that the skill differences between configurations are much larger than the sampling variability within an ensemble.

\foreach \p in {16,27,30,42,52} {
    \begin{figure}[h!]
        \centering
        \includegraphics[width=0.99\linewidth]{figures/DM/ensemble/cat_patch_\p.png}
        \caption{Ensemble-mean flood extent decomposed into hits, misses and false alarms at patch~\p, plotted as in Figure~\ref{fig:cat_8}.}
        \label{fig:cat_\p}
    \end{figure}
}

\subsection{Probabilistic flood maps}

\begin{figure}[h!]
    \centering
    \includegraphics[width=0.99\linewidth]{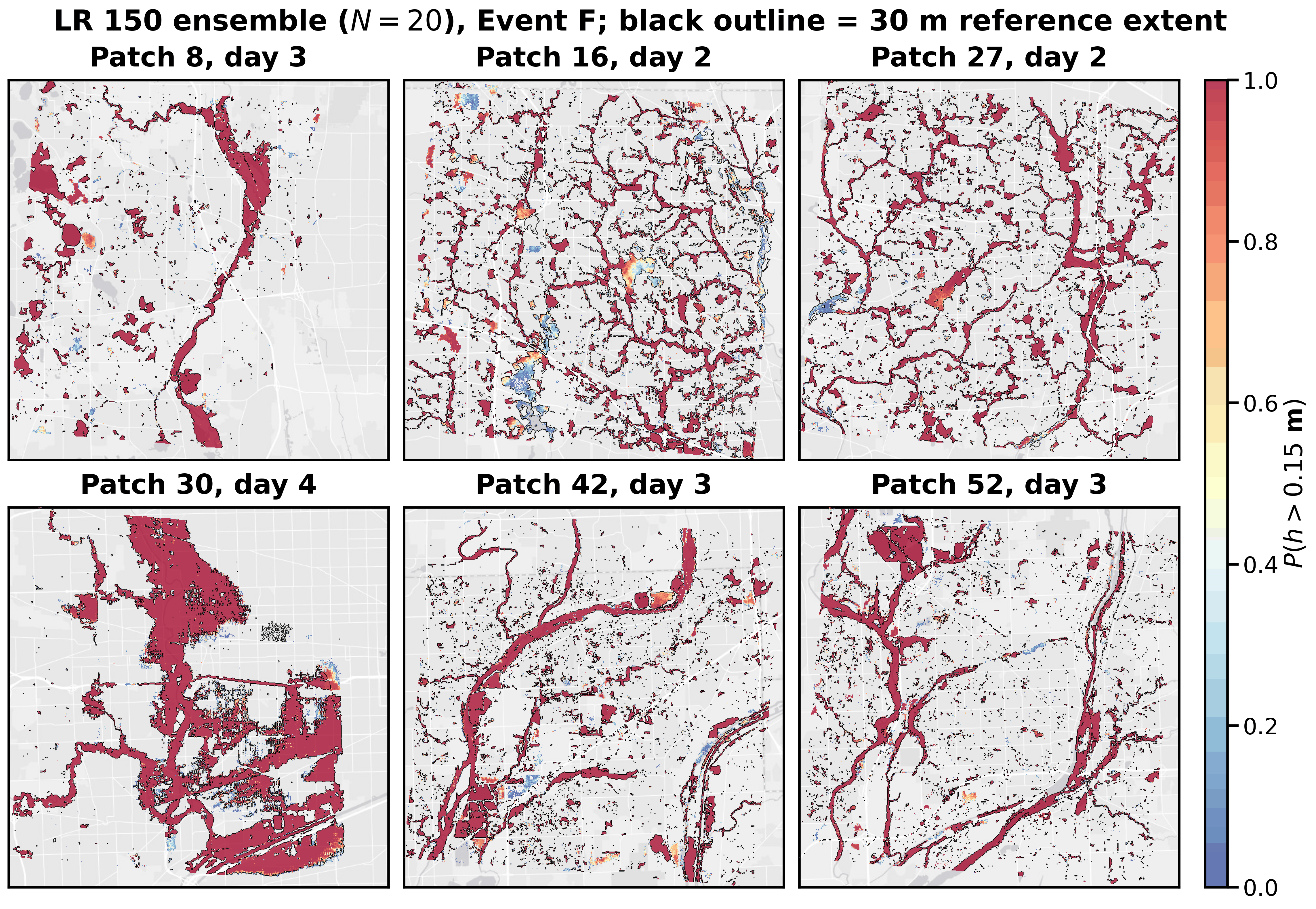}
    \caption{Exceedance probability $P(h > 0.15~\mathrm{m})$ from the 150~m conditioned 20-member ensemble for Event~F, at the day of maximum reference inundation, for the six analysis patches. The black outline is the 30~m reference flood extent. Probabilities below 0.02 are transparent.}
    \label{fig:prob_F}
\end{figure}

The product that distinguishes an ensemble from a deterministic forecast is the exceedance probability. Figure~\ref{fig:prob_F} maps $P(h > 0.15~\mathrm{m})$, the fraction of the 20 members exceeding the inundation threshold, for Event~F across all six analysis patches, with the reference flood outline overlaid.

The probability fields are spatially coherent and track the reference outline closely: cells within the channel and the inner floodplain carry probabilities near unity, cells on valley sides near zero. The informative structure is the transition band between them, which is where the conditioning leaves the position of the wetting front undetermined and where an operational user would concentrate attention.

That band is, however, narrow, typically only a few cells wide, so the probability field is close to binary. This sharpness is a direct visual expression of the under-dispersion quantified in Section~\ref{sec:verification}: the ensemble is confident, and more confident than its own error justifies. The maps are therefore useful for delineating the uncertain margin of the flood, but their probability values should not be read as calibrated exceedance frequencies.

\subsection{Aggregated performance}
\label{sec:agg}

\begin{figure}[h!]
    \centering
    \includegraphics[width=1\linewidth]{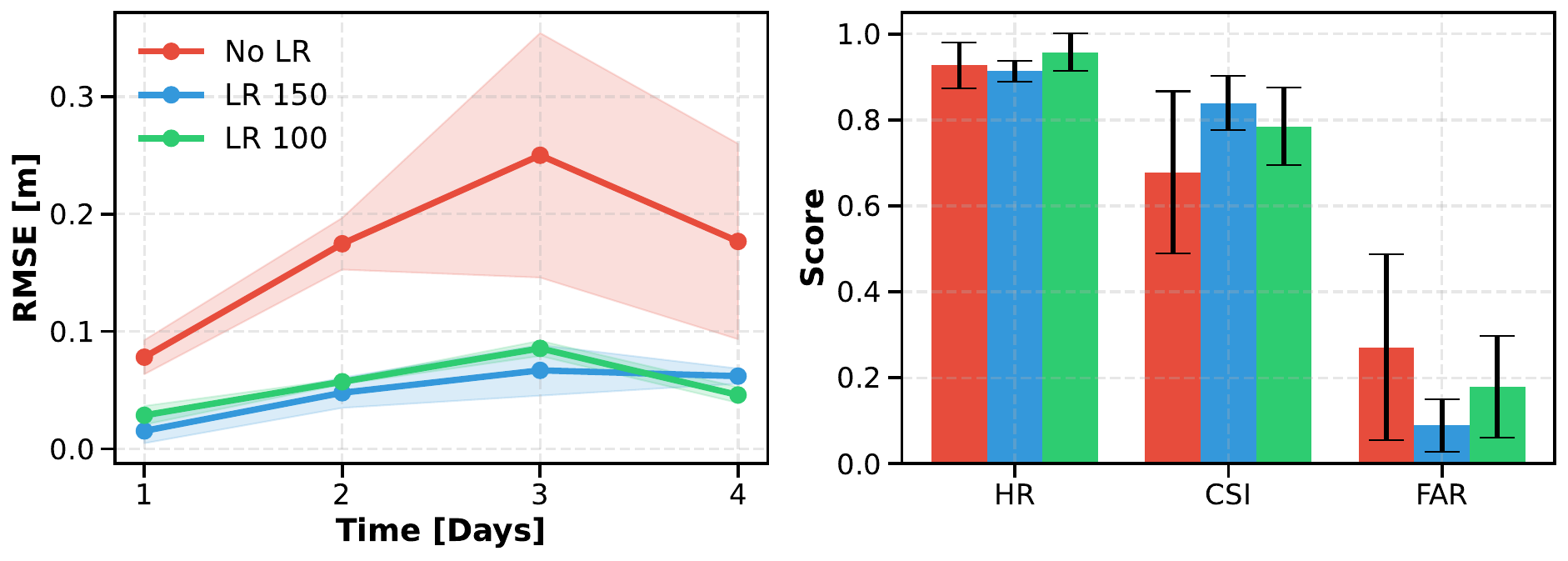}
    \caption{Aggregated ensemble performance over the six analysis patches and six testing events. Left: pooled RMSE of the ensemble mean against lead time, with shaded bands of one standard deviation across events. Right: pooled hit rate, critical success index and false alarm ratio of the ensemble mean on the day of maximum reference inundation, with error bars of one standard deviation across events.}
    \label{fig:agg_metrics}
\end{figure}

Figure~\ref{fig:agg_metrics} aggregates the ensemble-mean forecasts over the six analysis patches and six testing events. Squared errors and contingency counts are pooled across patches before the score is formed, so that larger inundated areas contribute in proportion to their extent. Table~\ref{tab:aggregate_metrics} lists the corresponding values.

The left panel shows the pooled RMSE of the ensemble mean against lead time. Both conditioned ensembles grow from a small day-1 error to a maximum on day~3 and recede thereafter, peaking at 0.067~m for the 150~m conditioning and 0.086~m for the 100~m conditioning. The unconditioned ensemble has the largest error at every lead time: its day-1 RMSE (0.078~m) already exceeds the day-3 peak of the 150~m configuration, and it peaks at 0.250~m on day~3. Its shaded band is three to five times wider, indicating that its error depends strongly on which storm is being forecast.

The right panel and Table~\ref{tab:aggregate_metrics} give the pooled peak-day contingency scores. The unconditioned ensemble is weakest on CSI (0.678) and FAR (0.271), and its four-day mean RMSE of 0.170~m is more than three times that of either conditioned configuration. Between the two conditioned configurations the pattern is consistent and is the main quantitative finding of this work: the 100~m conditioning attains the higher hit rate (0.957 against 0.913), but the 150~m conditioning is superior on the other three scores, with lower RMSE (0.048 against 0.054~m), higher CSI (0.839 against 0.785) and markedly lower false alarm ratio (0.089 against 0.179).

The mechanism is visible in the extent maps: the 100~m conditioned ensemble produces a systematically wetter forecast, which raises hits and false alarms together and nets out negative on the overlap-based CSI. The rank histograms of Section~\ref{sec:verification} confirm this bias independently. Because the 150~m conditioning is also the cheaper of the two to compute (Section~\ref{sec:cost}), a finer conditioning resolution does not buy better flood mapping within the range examined here, and the coarser option offers the better accuracy--cost trade-off.

As a check on the representativeness of the six analysis patches, these aggregate values were compared with single-realization forecasts over all 60 patches and all three antecedent scenarios, used only for this check. Those forecasts gave four-day mean RMSE of 0.190, 0.050 and 0.058~m and CSI of 0.689, 0.832 and 0.787 for the unconditioned, 150~m and 100~m configurations. Agreement to within roughly 0.02 in CSI and 0.02~m in RMSE indicates that the six analysis patches are representative of the basin as a whole, and that the ranking of the three configurations is not an artifact of the reduced evaluation scope.

\begin{figure}[h!]
    \centering
    \includegraphics[width=1\linewidth]{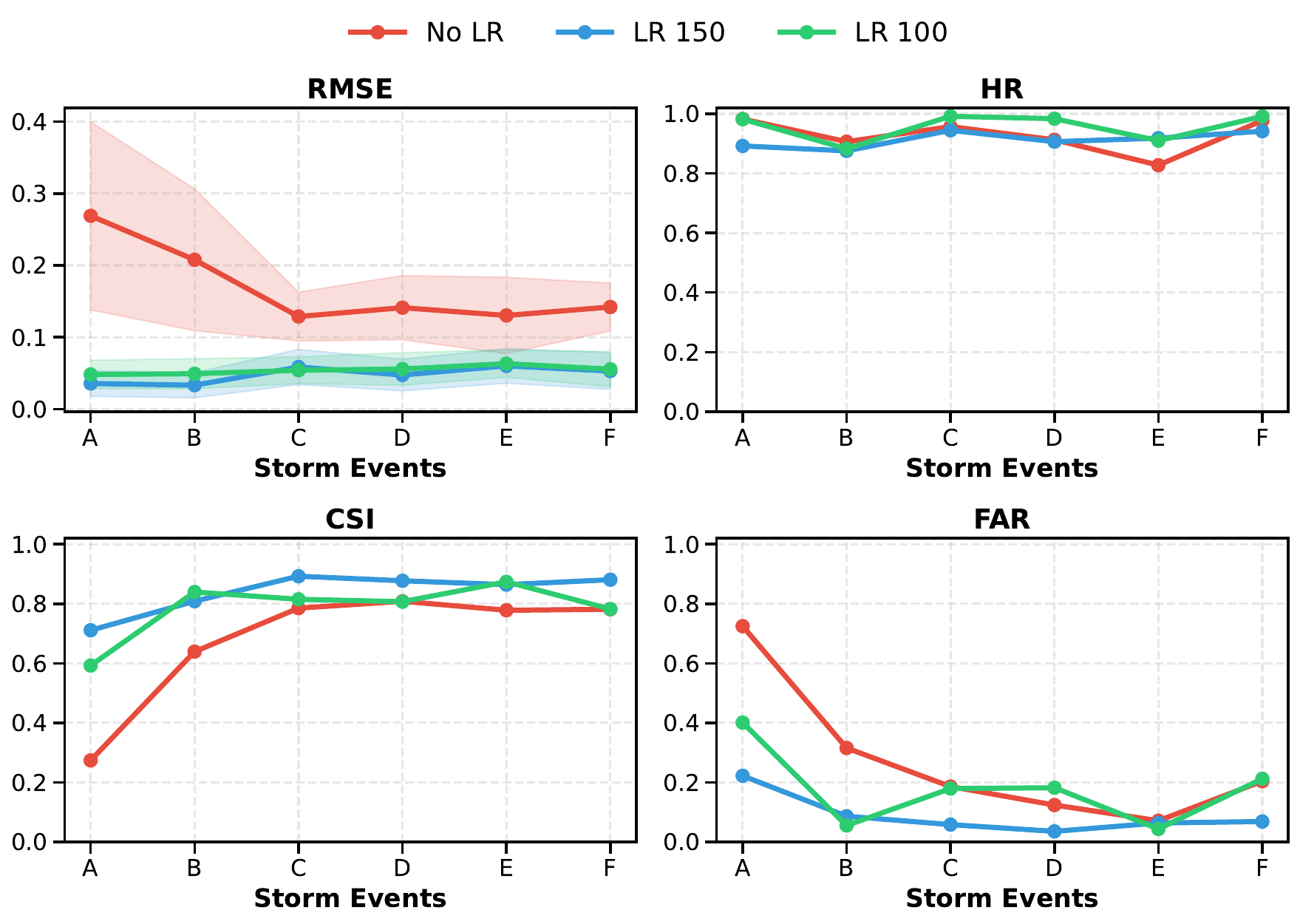}
    \caption{Aggregated ensemble performance resolved by testing event. RMSE is averaged over the four lead times with a shaded band of one standard deviation across lead times; HR, CSI and FAR are evaluated on the day of maximum reference inundation. The horizontal axis is categorical; see Table~\ref{tab:test_events} for the severity characteristics of each event.}
    \label{fig:agg_metrics_event}
\end{figure}

Figure~\ref{fig:agg_metrics_event} resolves the same four scores by testing event. The ordering of the configurations is stable across events for CSI and FAR. The unconditioned ensemble is worst for the weak Event~A by a wide margin and narrows the gap for the more severe events, consistent with the pattern noted in the extent maps: larger storms produce more spatially contiguous inundation that aligns with the geomorphic floodplain and HAND layers supplied as static conditioning, and such patterns are easier to reconstruct without hydrodynamic guidance. As noted in Section~\ref{sec:dataset_generation}, Events~A--F are six distinct severity regimes rather than a monotonic severity axis, so the horizontal axis should be read as categorical.

\begin{table}[!htb]
\centering
\caption{Aggregated ensemble-mean performance over the six analysis patches and six testing events at the 0.1~mm/hr antecedent scenario. RMSE is the pooled value averaged over the four forecast days; HR, CSI and FAR are pooled on the day of maximum reference inundation. Best value per metric in bold.}
\label{tab:aggregate_metrics}
\begin{tabular}{lcccc}
\toprule
\textbf{Configuration} & \textbf{RMSE [m]} $\downarrow$ & \textbf{HR} $\uparrow$ & \textbf{CSI} $\uparrow$ & \textbf{FAR} $\downarrow$ \\
\midrule
DDPM + LR (100 m) & 0.054          & \textbf{0.957} & 0.785          & 0.179 \\
DDPM + LR (150 m) & \textbf{0.048} & 0.913          & \textbf{0.839} & \textbf{0.089} \\
DDPM (No LR)      & 0.170          & 0.928          & 0.678          & 0.271 \\
\bottomrule
\end{tabular}
\end{table}

\subsection{Probabilistic verification and calibration}
\label{sec:verification}

The results above use the ensemble mean as a point forecast. This subsection assesses the ensemble \emph{distribution}, which determines whether the probabilistic products of Section~\ref{sec:ensemble_method} can be trusted. Four standard diagnostics are used, all computed over valid cells at each of the four lead times.

The \textbf{continuous ranked probability score} \cite{matheson1976scoring,hersbach2000decomposition} is a strictly proper scoring rule \cite{gneiting2007strictly} generalizing the mean absolute error to probabilistic forecasts. The fair (unbiased) ensemble estimator is used,
\begin{equation}
    \mathrm{CRPS} = \frac{1}{N}\sum_{i=1}^{N}\lvert x_i - y\rvert
    - \frac{1}{2N(N-1)}\sum_{i=1}^{N}\sum_{j=1}^{N}\lvert x_i - x_j\rvert ,
    \label{eq:crps}
\end{equation}
with $x_i$ the members and $y$ the reference. \textbf{Spread--skill consistency} compares the root mean ensemble variance with the RMSE of the ensemble mean, which should match for a statistically reliable ensemble \cite{fortin2014why}. The \textbf{rank histogram} \cite{talagrand1997evaluation,hamill2001interpretation} accumulates the rank of the reference value within the sorted ensemble over cells that are wet in the reference; a flat histogram indicates calibration, a U shape under-dispersion, and an asymmetric spike a systematic bias. The \textbf{Brier score} \cite{brier1950verification} and a reliability diagram assess the exceedance probability.

\begin{figure}[h!]
    \centering
    \includegraphics[width=1\linewidth]{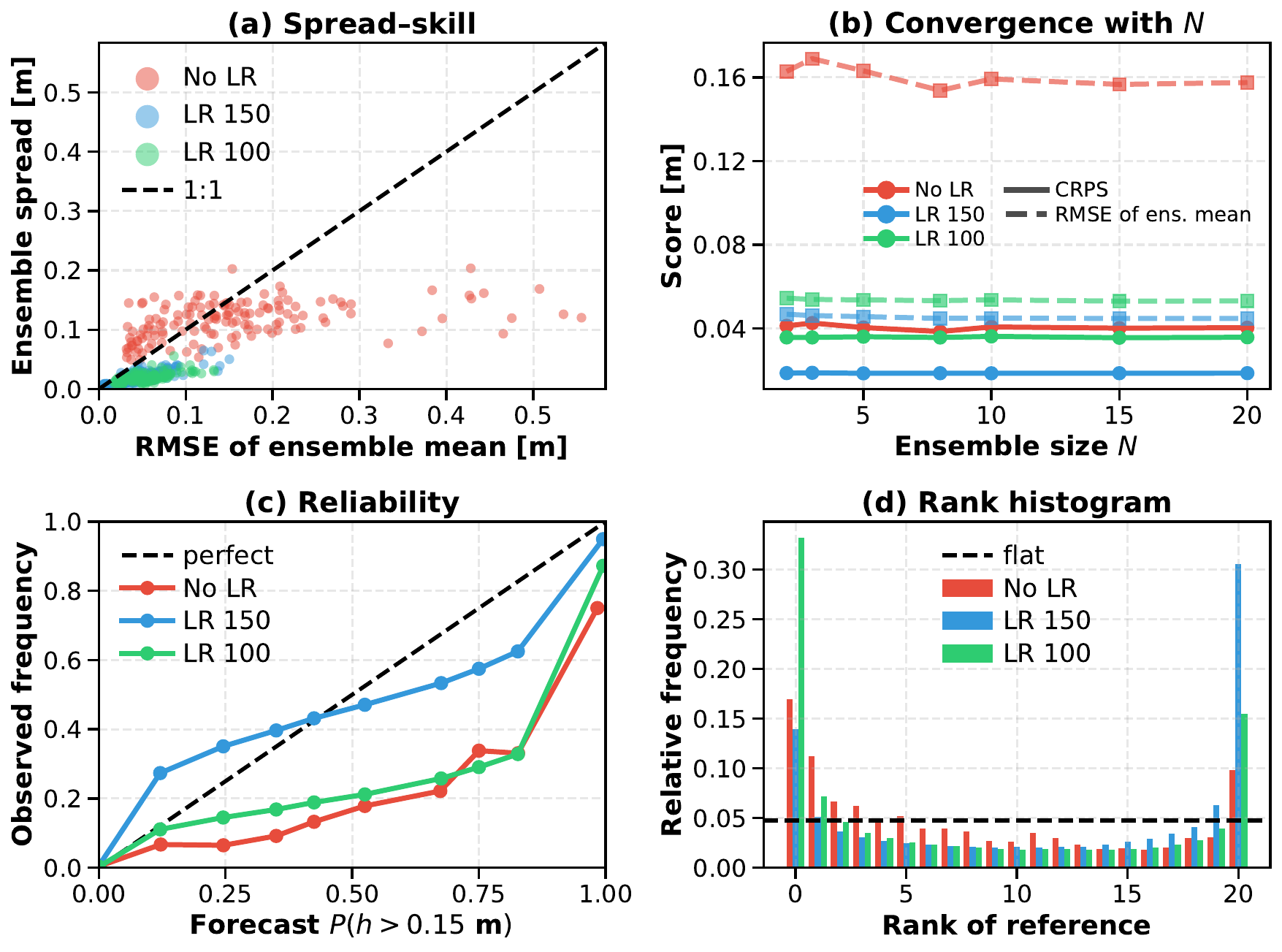}
    \caption{Probabilistic verification of the DDPM ensembles. (a) Spread--skill relationship, one point per case and lead time, with the 1:1 line indicating statistical consistency; all configurations lie below it. (b) Convergence of CRPS (solid) and ensemble-mean RMSE (dashed) with ensemble size $N$. (c) Reliability diagram for the exceedance probability $P(h>0.15~\mathrm{m})$. (d) Rank histogram of the reference value within the ensemble over cells wet in the reference; the dashed line is the flat histogram expected of a calibrated ensemble.}
    \label{fig:ensemble_scores}
\end{figure}

Table~\ref{tab:ensemble_metrics} and Figure~\ref{fig:ensemble_scores} report the results. Unlike the pooled scores of Section~\ref{sec:agg}, each score here is computed separately for every patch, event and lead time and then averaged with equal weight, so the ensemble-mean RMSE values differ slightly from those of Table~\ref{tab:aggregate_metrics} (e.g.\ 0.045 versus 0.048~m for the 150~m configuration) without changing the ranking. Three findings follow.

First, averaging the members is worthwhile, and increasingly so as the conditioning weakens. The ensemble mean reduces water-depth RMSE relative to the mean of the individual members by 4.9\%, 8.2\% and 10.7\% for the 100~m, 150~m and unconditioned configurations. This ordering is expected, since the unconditioned model has the most freedom and therefore the largest share of single-realization error that averaging can remove.

Second, the probabilistic ranking of the configurations reproduces the point-forecast ranking, and more emphatically. The 150~m conditioned ensemble attains the best value on every probabilistic score: CRPS of 0.019~m against 0.036~m at 100~m and 0.040~m unconditioned, and Brier score of 0.014 against 0.019 and 0.045. The CRPS separation between the two conditioned configurations is proportionally much larger than their separation in ensemble-mean RMSE, indicating that the advantage of the coarser conditioning lies in the shape of its predictive distribution and not only in the accuracy of its central estimate.

Third, and essential for interpreting the probability maps, all three ensembles are substantially \textbf{under-dispersive}. The ratio of ensemble spread to ensemble-mean error is 0.35, 0.45 and 0.73 for the 100~m, 150~m and unconditioned configurations, against unity for a statistically consistent ensemble. The ensembles are overconfident, and most so where the conditioning is most informative. The rank histograms in Figure~\ref{fig:ensemble_scores}d are strongly U-shaped in every case, with diagnostic asymmetries: the 100~m configuration places 33\% of reference values in the lowest rank, meaning the reference depth falls below every member and the forecast is wet-biased, whereas the 150~m configuration places 31\% in the highest rank and is dry-biased. These two signatures independently confirm the contrasting hit-rate and false-alarm behavior of Section~\ref{sec:agg}, recovered from the distributional structure of the forecast rather than from thresholded extent. The reliability diagram in Figure~\ref{fig:ensemble_scores}c agrees, with the 100~m and unconditioned curves below the diagonal over most of the probability range and the 150~m curve closest to it.

A practical consequence follows from Figure~\ref{fig:ensemble_scores}b. Because the generative spread is narrow, CRPS and ensemble-mean RMSE converge by roughly three to five members and are flat thereafter; increasing $N$ from 5 to 20 changes CRPS by less than 2\% for every configuration. Large ensembles are therefore unnecessary for this system as trained, which substantially reduces the cost of the probabilistic product. The deeper implication is that resampling the diffusion noise is not a complete uncertainty budget: the spread it generates is real but narrow, and an operational system would need to perturb the meteorological forcing as well, through the nested design described in Section~\ref{sec:ensemble_method}. Calibrating the generative component directly, for instance by post-hoc spread inflation or by training with a dispersion-aware objective, is a natural extension not attempted here.

\begin{table}[!htb]
\centering
\caption{Probabilistic verification of the 20-member ensembles over the six analysis patches and six testing events, computed per patch, event and lead time and averaged with equal weight (hence the small differences from the pooled RMSE of Table~\ref{tab:aggregate_metrics}). Water depth in meters. A spread-to-error ratio below unity indicates an under-dispersive (overconfident) ensemble. Best value per column in bold.}
\label{tab:ensemble_metrics}
\begin{tabular}{lccccc}
\toprule
\textbf{Configuration} & \textbf{RMSE} & \textbf{RMSE} & \textbf{CRPS} $\downarrow$ & \textbf{Spread/} & \textbf{Brier} $\downarrow$ \\
 & \textbf{single member} & \textbf{ens.\ mean} $\downarrow$ & & \textbf{error ratio} & \\
\midrule
DDPM + LR (100 m) & 0.056 & 0.053 & 0.036 & 0.35 & 0.019 \\
DDPM + LR (150 m) & \textbf{0.049} & \textbf{0.045} & \textbf{0.019} & 0.45 & \textbf{0.014} \\
DDPM (No LR)      & 0.176 & 0.158 & 0.040 & \textbf{0.73} & 0.045 \\
\bottomrule
\end{tabular}
\end{table}

\begin{figure}[h!]
    \centering
    \includegraphics[width=1\linewidth]{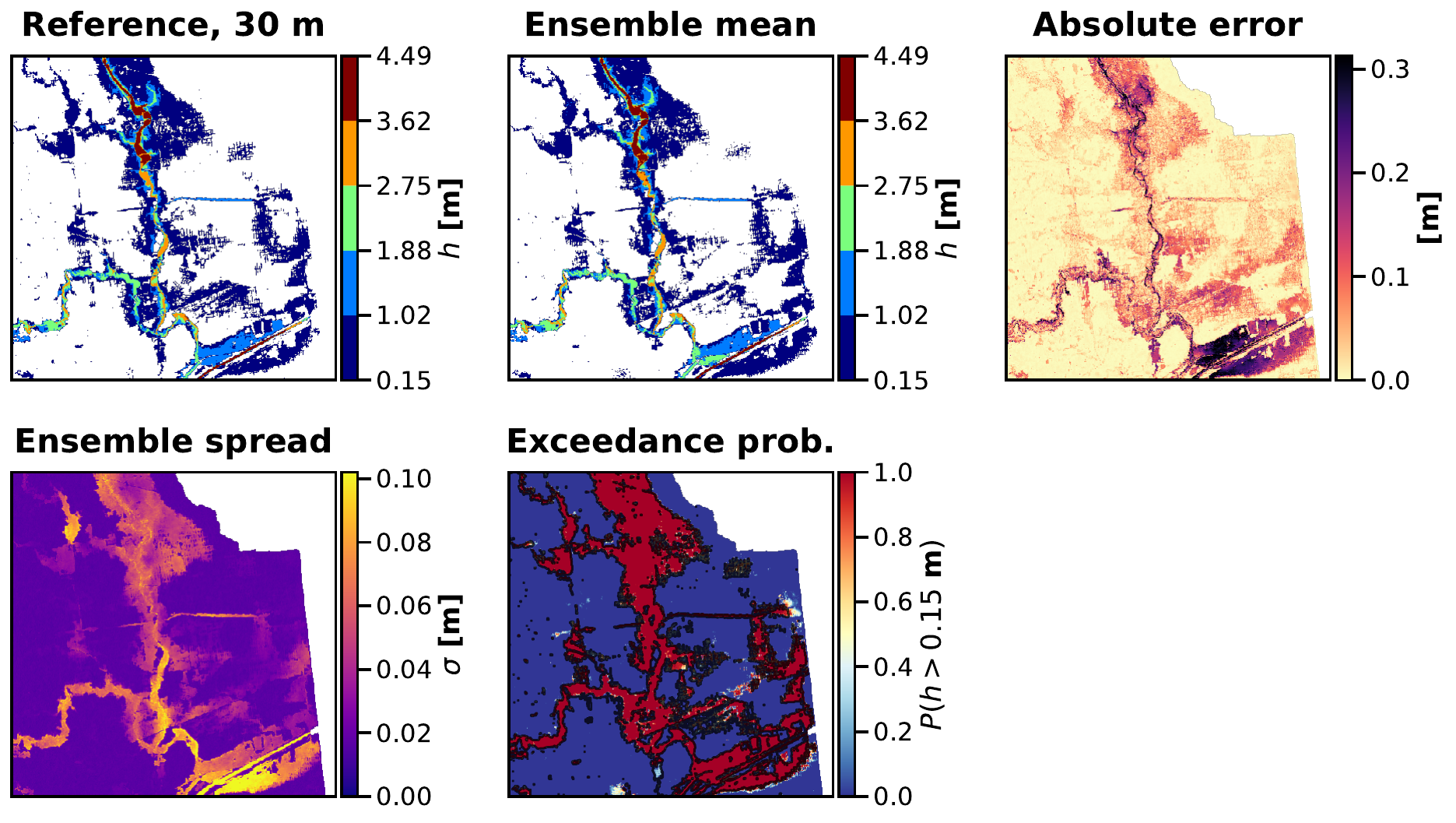}
    \caption{Anatomy of a single ensemble forecast: Event~F at patch~30 and a lead time of three days, 150~m conditioning, $N=20$. From left to right: the 30~m SynxFlow reference water depth; the ensemble mean; the absolute error of the ensemble mean; the ensemble standard deviation; and the exceedance probability $P(h>0.15~\mathrm{m})$ with the reference flood outline in black. The spread concentrates along the margins of the inundated area, where the conditioning leaves the position of the wetting front undetermined.}
    \label{fig:ensemble_maps}
\end{figure}

\subsection{Computational cost}
\label{sec:cost}

The computational cost of the framework was evaluated for a complete four-day forecast over the Des Plaines River watershed, as summarized in Table~\ref{tab:speedup}. All timings were measured on NVIDIA A100 GPUs of the same cluster. For operational inference, the watershed is tiled with 26 non-overlapping $512\times512$ patches, and the DDPM cost is obtained from the measured per-patch inference time.

The hydrodynamic simulations used to build the training dataset were executed on eight GPUs; averaged over the 339 archived simulations, a 121-hour run required 538~s at 30~m, 76~s at 100~m and 43.6~s at 150~m of wall-clock time. Because the DDPM performs inference on a single GPU, a dedicated single-GPU benchmark of the solver was additionally run for testing Event~F under the driest antecedent scenario, giving the like-for-like single-device costs in Table~\ref{tab:speedup}. A single-member watershed-scale four-day forecast, comprising 26 patch evaluations at 100 denoising steps, costs 120~s on one GPU for the conditioned configurations and 122~s unconditioned, and is essentially independent of the conditioning configuration (Table~\ref{tab:time_tuning}).

Relative to the single-GPU 30~m reference simulation, a single-member forecast achieves speedups of $7.8\times$ and $11.0\times$ when conditioned on 100~m and 150~m simulations, since the coarse simulation must still be run first, and $16.3\times$ without hydrodynamic conditioning.

The advantage is larger for the ensemble forecasts reported in this paper, because the conditioning inputs, including any coarse hydrodynamic simulation, are computed once and reused by every member. The measured marginal cost of an additional member is 3.95~s per patch, so a watershed-scale $N$-member ensemble costs $103N$~s. Writing $T_{30}$ for one 30~m simulation, $T_{\mathrm{LR}}$ for one coarse simulation and $T_{\mathrm{DM}} = 103$~s per member, the speedup of an $N$-member ensemble over an $N$-member physics-based ensemble is
\begin{equation}
    S(N) = \frac{N\,T_{30}}{T_{\mathrm{LR}} + N\,T_{\mathrm{DM}}},
    \label{eq:ensemble_speedup}
\end{equation}
which increases monotonically with $N$ towards $T_{30}/T_{\mathrm{DM}} = 19.3\times$. Table~\ref{tab:ensemble_cost} evaluates it. Because $T_{\mathrm{DM}}$ is the marginal cost of a member sampled in batches of ten, Eq.~\eqref{eq:ensemble_speedup} at $N=1$ gives slightly higher values ($8.4\times$, $12.2\times$ and $19.3\times$) than the unbatched single-member speedups of Table~\ref{tab:speedup} ($7.8\times$, $11.0\times$ and $16.3\times$); the latter apply to a single forecast. A 20-member watershed ensemble costs about 0.6~GPU-hours against 11.0~GPU-hours for the equivalent set of 30~m simulations. Accelerated or distilled samplers, which reduce the number of reverse diffusion steps per member, offer a direct route to further reductions.

Beyond these measured ratios, the surrogate cost scales favorably: inference grows weakly with forecast horizon because lead times are emitted as output channels rather than integrated sequentially under a CFL constraint; it scales with forecast area through independent patch evaluations, whereas the 30~m SWE solution becomes rapidly more expensive as the domain grows; and, as Eq.~\eqref{eq:ensemble_speedup} shows, it improves with ensemble size.

\paragraph{Projected speedup.} The forecast horizon can be extended to 14 days by adding output channels, so a single network pass still produces the full forecast while the SWE runtime grows in proportion to the simulated period (by a factor of $361/121 \approx 3$). Table~\ref{tab:projected_speedup} reports the resulting speedups relative to the single-GPU 30~m simulation; these are projections from the measured timings of Table~\ref{tab:speedup}.

\begin{table}[!htb]
\centering
\caption{Computational cost of a four-day flood forecast over the Des Plaines River watershed. Hydrodynamic timings are for a 121-hour SynxFlow run; the eight-GPU column is the mean over the 339 archived dataset simulations and the single-GPU column is a dedicated benchmark for testing Event~F under the 0.1~mm/hr antecedent scenario. For that event the matched eight-GPU runtimes were 487, 67 and 38~s at 30, 100 and 150~m, so the single-GPU benchmark corresponds to parallel speedups of $4.1\times$, $2.0\times$ and $1.6\times$ on eight devices. DDPM inference covers the 26 non-overlapping patches at 100 denoising steps on one GPU, for one ensemble member. Speedups are referenced to the single-GPU 30~m simulation.}
\label{tab:speedup}
\begin{tabular}{lccc}
\toprule
\textbf{Configuration} & \textbf{8 GPUs [s]} & \textbf{1 GPU [s]} & \textbf{Speedup vs.\ 30~m (1 GPU)} \\
\midrule
SynxFlow 30 m (reference)   & 538  & 1986 & -- \\
SynxFlow 100 m              & 76   & 133  & -- \\
SynxFlow 150 m              & 44   & 60   & -- \\
DDPM, one member            & --   & 120--122 & -- \\
\midrule
DDPM + LR (100 m)           & --   & 253 & $7.8\times$ \\
DDPM + LR (150 m)           & --   & 180 & $11.0\times$ \\
DDPM (No LR)                & --   & 122 & $16.3\times$ \\
\bottomrule
\end{tabular}
\end{table}

\begin{table}[!htb]
\centering
\caption{Speedup of an $N$-member watershed-scale ensemble on a single GPU, from Eq.~\eqref{eq:ensemble_speedup} with the timings of Table~\ref{tab:speedup} and $T_{\mathrm{DM}} = 103$~s per member, the marginal cost of a member sampled in batches of ten. The physics-based reference generates $N$ independent 30~m simulations. The $N=1$ column is the batched-cost limit; the unbatched single-member speedups are given in Table~\ref{tab:speedup}.}
\label{tab:ensemble_cost}
\begin{tabular}{lccc}
\toprule
\textbf{Configuration} & $N=1$ & $N=5$ & $N=20$ \\
\midrule
DDPM + LR (100 m)  & $8.4\times$  & $15.4\times$ & $18.2\times$ \\
DDPM + LR (150 m)  & $12.2\times$ & $17.3\times$ & $18.8\times$ \\
DDPM (No LR)       & $19.3\times$ & $19.3\times$ & $19.3\times$ \\
\bottomrule
\end{tabular}
\end{table}

\begin{table}[!htb]
\centering
\caption{Projected speedup relative to the single-GPU 30~m SynxFlow simulation for a 14-day forecast with 26 non-overlapping patches.}
\label{tab:projected_speedup}
\begin{tabular}{lcc}
\toprule
\textbf{Configuration} & \textbf{1 member} & \textbf{20 members} \\
\midrule
DDPM + LR (100 m) & $11.5\times$ & $48.4\times$ \\
DDPM + LR (150 m) & $19.8\times$ & $53.1\times$ \\
DDPM (No LR)      & $48.7\times$ & $57.7\times$ \\
\bottomrule
\end{tabular}
\end{table}

\clearpage
\newpage
\section{Conclusions}
\label{sec:conclusion}

In this study, a conditional denoising diffusion probabilistic model (DDPM) was developed for two-dimensional spatiotemporal flood inundation forecasting within the Des Plaines River watershed in northeastern Illinois. The model predicts the evolution of 30~m water depth at four daily lead times in a single network evaluation, rather than a binary inundation extent or a peak-flood snapshot, and therefore captures both the propagation and the recession of the flood. It was trained on a physically based dataset generated with the SynxFlow multi-GPU shallow-water-equation solver, driven by 113 historical precipitation events observed between 2003 and 2024 across twelve states of the Midwestern United States and simulated under three antecedent-wetness scenarios, giving 339 hydrodynamic simulations per grid resolution that span a wide range of storm intensities, precipitation patterns and antecedent conditions.

Conditioning on low-resolution hydrodynamic simulations substantially improved the accuracy and temporal consistency of the forecasts. It reduced the four-day mean water-depth RMSE of the ensemble mean from 0.170~m for the unconditioned model to 0.048~m and 0.054~m for the 150~m and 100~m conditioned configurations, respectively, and raised the critical success index from 0.678 to 0.839 and 0.785. The conditioned models generalized well to the six withheld testing storms, including the event with the most extreme short-duration rainfall intensity, showing that the coarse hydrodynamic solution supplies the temporal structure of the flood while the diffusion model supplies the fine-scale spatial detail that the coarse simulation cannot resolve.

A practically useful finding is that the coarser 150~m conditioning outperformed the finer 100~m conditioning on water-depth RMSE, critical success index, false alarm ratio, CRPS and Brier score, while also being the less expensive option. The 100~m conditioned model achieved a higher hit rate, but did so by producing a systematically wetter forecast that also increased its false alarm ratio. Within the range examined, a finer conditioning resolution therefore does not guarantee better flood-mapping skill, and the coarser option offers the better accuracy--cost trade-off.

Because the reverse diffusion process is a stochastic Markov chain, the trained model is probabilistic by construction, and every forecast in this study is a 20-member ensemble obtained without retraining, input perturbation or architectural modification. Averaging the members reduced water-depth RMSE relative to an individual realization by 4.9\%, 8.2\% and 10.7\% for the 100~m, 150~m and unconditioned configurations, and the ensemble spread provided spatially explicit flood exceedance probabilities. The 150~m configuration was again the best performer, attaining the lowest CRPS and Brier score. Probabilistic verification further showed that the ensembles are under-dispersive, with spread-to-error ratios of 0.35--0.73, and their rank histograms identify a wet bias at 100~m and a dry bias at 150~m that is consistent with the hit-rate and false-alarm behavior of the flood-extent analysis. Because the generative spread is narrow, the scores converge with three to five members, so compact ensembles are sufficient, and combining the generative ensemble with perturbed meteorological forcing is the natural route to a complete operational uncertainty estimate.

From a computational perspective, a single-member four-day forecast on one GPU was $7.8\times$, $11.0\times$ and $16.3\times$ faster than the 30~m hydrodynamic simulation for the 100~m, 150~m and unconditioned configurations. The advantage increases with ensemble size, because the conditioning inputs, including any coarse hydrodynamic simulation, are computed once and reused by every member, and with forecast horizon, because additional lead days are emitted as output channels rather than integrated in time. For a 14-day forecast, the projected speedups reach $11.5\times$, $19.8\times$ and $48.7\times$ for a single member and $48.4\times$, $53.1\times$ and $57.7\times$ for a 20-member ensemble.

Overall, this work demonstrates that physics-guided diffusion models can serve as efficient and accurate probabilistic surrogates for regional-scale flood forecasting. By combining the physical consistency of hydrodynamic simulations with the generative capability of diffusion models, the framework preserves high-resolution inundation structure at reduced computational cost while natively providing the ensemble information that operational flood warning requires, and its patch-based formulation provides a scalable pathway to additional watersheds. Future work will extend the framework toward regional-scale flood forecasting across Illinois, replace the domain-wide precipitation descriptor with patch-local forcing, drive the model with numerical weather prediction ensembles so that meteorological and generative uncertainty are propagated jointly, integrate riverine boundary conditions, and explore accelerated samplers to further reduce the cost of large ensembles.

\section*{CRediT authorship contribution statement}
\textbf{Omar Sallam:} Conceptualization, Methodology, Software, Formal analysis, Investigation, Data curation, Visualization, Writing -- original draft.
\textbf{David Coe:} Writing -- review \& editing.
\textbf{Jeremy Feinstein:} Writing -- review \& editing.
\textbf{Hugo Torres:} Writing -- review \& editing.
\textbf{Sofia Sahagun Covarrubias:} Writing -- review \& editing.
\textbf{Eugene Yan:} Supervision, Funding acquisition, Project administration, Writing -- review \& editing.

\section*{Declaration of competing interest}
The authors declare that they have no known competing financial interests or personal relationships that could have appeared to influence the work reported in this paper.

\section*{Acknowledgment}
This research used resources of the Laboratory Computing Resource Center (LCRC) at Argonne National Laboratory. Argonne National Laboratory is operated by UChicago Argonne, LLC, for the U.S. Department of Energy under Contract No.~DE-AC02-06CH11357.

\section*{Data availability}
The precipitation forcing is derived from the publicly available NCEP/NWS Stage~IV multisensor precipitation analysis. Terrain and land-cover data are from the USGS 3DEP, USGS NHD, and NLCD~2021 public archives. The SynxFlow solver is open source. The generated hydrodynamic simulation dataset, the trained model weights, and the analysis scripts are available from the corresponding author on reasonable request.

\clearpage
\newpage
\bibliographystyle{elsarticle-num}
\bibliography{references}

\begin{thebibliography}{10}
\expandafter\ifx\csname url\endcsname\relax
  \def\url#1{\texttt{#1}}\fi
\expandafter\ifx\csname urlprefix\endcsname\relax\def\urlprefix{URL }\fi
\expandafter\ifx\csname href\endcsname\relax
  \def\href#1#2{#2} \def\path#1{#1}\fi

\bibitem{CRED2025Disasters}
{Centre for Research on the Epidemiology of Disasters (CRED)},
  \href{https://reliefweb.int/report/world/2025-disasters-numbers}{2025
  disasters in numbers: Earth, wind and fire}, Tech. rep., Centre for Research
  on the Epidemiology of Disasters (CRED), accessed: 2026-05-29 (Apr. 2026).
\newline\urlprefix\url{https://reliefweb.int/report/world/2025-disasters-numbers}

\bibitem{seneviratne2021weather}
S.~I. Seneviratne, X.~Zhang, M.~Adnan, W.~Badi, C.~Dereczynski, A.~D. Luca,
  S.~Ghosh, I.~Iskandar, J.~Kossin, S.~Lewis, et~al., Weather and climate
  extreme events in a changing climate, Climate change 2021: The physical
  science basis: Working group I contribution to the sixth assessment report of
  the intergovernmental panel on climate change (2021) 1513--1766.

\bibitem{rogers2025role}
J.~S. Rogers, M.~P. Maneta, S.~R. Sain, L.~E. Madaus, J.~P. Hacker, The role of
  climate and population change in global flood exposure and vulnerability,
  Nature Communications 16~(1) (2025) 1287.
\newblock \href {https://doi.org/10.1038/s41467-025-56654-8}
  {\path{doi:10.1038/s41467-025-56654-8}}.

\bibitem{chen2026global}
W.~Chen, Y.~Zheng, Y.~Zhou, T.~Zhang, Y.~Zhang, D.~Z. Zhu, X.~Zhou, L.~Zhang,
  X.~Zhang, Z.~Huang, Global acceleration of compound flood risks through
  fluvial-tidal interactions in a warming climate, npj Natural Hazards 3~(1)
  (2026) 13.
\newblock \href {https://doi.org/10.1038/s44304-026-00179-7}
  {\path{doi:10.1038/s44304-026-00179-7}}.

\bibitem{ipcc2022wg2}
{IPCC}, Climate change 2022: Impacts, adaptation and vulnerability, Tech. rep.,
  Intergovernmental Panel on Climate Change (2022).

\bibitem{wmo2022earlywarnings}
{World Meteorological Organization},
  \href{https://public.wmo.int/en/resources/bulletin/early-warnings-all}{Early
  warnings for all: Executive action plan 2023--2027} (2022).
\newline\urlprefix\url{https://public.wmo.int/en/resources/bulletin/early-warnings-all}

\bibitem{wmo_floodforecasting}
{World Meteorological Organization}, Manual on flood forecasting and warning
  (2011).

\bibitem{teng2017flood}
J.~Teng, A.~J. Jakeman, J.~Vaze, B.~F. Croke, D.~Dutta, S.~Kim, Flood
  inundation modelling: A review of methods, recent advances and uncertainty
  analysis, Environmental modelling \& software 90 (2017) 201--216.
\newblock \href {https://doi.org/10.1016/j.envsoft.2017.01.006}
  {\path{doi:10.1016/j.envsoft.2017.01.006}}.

\bibitem{cloke2009ensemble}
H.~L. Cloke, F.~Pappenberger, Ensemble flood forecasting: A review, Journal of
  Hydrology 375~(3-4) (2009) 613--626.

\bibitem{asif2025review}
M.~Asif, M.~M. Kuglitsch, I.~Pelivan, R.~Albano, Review and intercomparison of
  machine learning applications for short-term flood forecasting, Water
  Resources Management 39~(5) (2025) 1971--1991.
\newblock \href {https://doi.org/10.1007/s11269-025-04093-x}
  {\path{doi:10.1007/s11269-025-04093-x}}.

\bibitem{liu2025comprehensive}
B.~Liu, Y.~Li, M.~Ma, B.~Mao, A comprehensive review of machine learning
  approaches for flood depth estimation, International Journal of Disaster Risk
  Science 16~(3) (2025) 433--445.
\newblock \href {https://doi.org/10.1007/s13753-025-00639-0}
  {\path{doi:10.1007/s13753-025-00639-0}}.

\bibitem{zahura2020training}
F.~T. Zahura, J.~L. Goodall, J.~M. Sadler, Y.~Shen, M.~M. Morsy, M.~Behl,
  Training machine learning surrogate models from a high-fidelity physics-based
  model: Application for real-time street-scale flood prediction in an urban
  coastal community, Water Resources Research 56~(10) (2020) e2019WR027038.

\bibitem{ma2021xgboost}
M.~Ma, G.~Zhao, B.~He, Q.~Li, H.~Dong, S.~Wang, Z.~Wang, Xgboost-based method
  for flash flood risk assessment, Journal of Hydrology 598 (2021) 126382.

\bibitem{lee2024predicting}
C.-C. Lee, L.~Huang, F.~Antolini, M.~Garcia, A.~Juan, S.~D. Brody,
  A.~Mostafavi, Predicting peak inundation depths with a physics informed
  machine learning model, Scientific Reports 14~(1) (2024) 14826.
\newblock \href {https://doi.org/10.1038/s41598-024-65570-8}
  {\path{doi:10.1038/s41598-024-65570-8}}.

\bibitem{liu2025pinn}
W.~Liu, Q.~Wu, Z.~Liu, X.~Zheng, T.~Liu, H.~Gao, Pinn framework for urban flood
  depth prediction: integrating data-driven insights with physical constraints,
  Journal of Hydrology 665 (2026) 134693.
\newblock \href {https://doi.org/10.1016/j.jhydrol.2025.134693}
  {\path{doi:10.1016/j.jhydrol.2025.134693}}.

\bibitem{sun2023rapid}
A.~Y. Sun, Z.~Li, W.~Lee, Q.~Huang, B.~R. Scanlon, C.~Dawson, Rapid flood
  inundation forecast using fourier neural operator, in: Proceedings of the
  IEEE/CVF international conference on computer Vision, 2023, pp. 3733--3739.

\bibitem{xu2024large}
Q.~Xu, Y.~Shi, J.~L. Bamber, C.~Ouyang, X.~X. Zhu, Large-scale flood modeling
  and forecasting with floodcast, Water Research 264 (2024) 122162.

\bibitem{xu2025floodcastbench}
Q.~Xu, Y.~Shi, J.~Zhao, X.~X. Zhu, Floodcastbench: A large-scale dataset and
  foundation models for flood modeling and forecasting, Scientific Data 12~(1)
  (2025) 431.

\bibitem{kumar2025temporal}
C.~Kumar, D.~McSpadden, M.~Schram, H.~Richter, Y.~Wang, B.~Roy, J.~L. Goodall,
  \href{https://s3.us-east-1.amazonaws.com/climate-change-ai/papers/neurips2025/26/slides.pdf}{Ai-driven
  temporal super-resolution for flooding prediction in norfolk, virginia},
  Presented at the NeurIPS 2025 Workshop on Tackling Climate Change with
  Machine Learning, climate Change AI (CCAI) Workshop, San Diego, CA, USA
  (2025).
\newline\urlprefix\url{https://s3.us-east-1.amazonaws.com/climate-change-ai/papers/neurips2025/26/slides.pdf}

\bibitem{si2026toward}
P.~Si, Y.~Qiu, O.~Sallam, J.~Feinstein, Z.~He, E.~Yan, P.~Chen, Toward
  ai-driven digital twins for metropolitan floods: A conditional latent
  dynamics network surrogate of the shallow water equations, Journal of
  Hydrology 680 (2026) 136461.
\newblock \href {https://doi.org/10.1016/j.jhydrol.2026.136461}
  {\path{doi:10.1016/j.jhydrol.2026.136461}}.

\bibitem{bentivoglio2023rapid}
R.~Bentivoglio, E.~Isufi, S.~N. Jonkman, R.~Taormina, Rapid spatio-temporal
  flood modelling via hydraulics-based graph neural networks, Hydrology and
  Earth System Sciences 27~(23) (2023) 4227--4246.

\bibitem{kazadi2024floodgnn}
A.~Kazadi, J.~Doss-Gollin, A.~Sebastian, A.~Silva, Floodgnn-gru: a
  spatio-temporal graph neural network for flood prediction, Environmental Data
  Science 3 (2024) e21.
\newblock \href {https://doi.org/10.1017/eds.2024.19}
  {\path{doi:10.1017/eds.2024.19}}.

\bibitem{kazadi2024pluvial}
A.~Kazadi, J.~Doss-Gollin, A.~L. Da~Silva, Pluvial flood emulation with
  hydraulics-informed message passing, in: Forty-First International Conference
  on Machine Learning, 2024.

\bibitem{bentivoglio2025multi}
R.~Bentivoglio, E.~Isufi, S.~N. Jonkman, R.~Taormina, Multi-scale hydraulic
  graph neural networks for flood modelling, Natural Hazards and Earth System
  Sciences 25~(1) (2025) 335--351.

\bibitem{taghizadeh2025interpretable}
M.~Taghizadeh, Z.~Zandsalimi, M.~A. Nabian, M.~Shafiee-Jood, N.~Alemazkoor,
  Interpretable physics-informed graph neural networks for flood forecasting,
  Computer-Aided Civil and Infrastructure Engineering 40~(18) (2025)
  2629--2649.
\newblock \href {https://doi.org/10.1111/mice.13484}
  {\path{doi:10.1111/mice.13484}}.

\bibitem{acosta2025physics}
C.~M. Acosta, H.~M. V.~V. Herath, J.~Y. Lim, A.~Saha, S.~Rasnayaka,
  L.~Marshall, {DUALFloodGNN}: Physics-informed graph neural network for
  operational flood modeling, in: Proceedings of the Thirty-Fifth International
  Joint Conference on Artificial Intelligence (IJCAI-26), 2026, pp. 6408--6416.
\newblock \href {https://doi.org/10.24963/ijcai.2026/713}
  {\path{doi:10.24963/ijcai.2026/713}}.

\bibitem{guo2022data}
Z.~Guo, V.~Moosavi, J.~P. Leit{\~a}o, Data-driven rapid flood prediction
  mapping with catchment generalizability, Journal of Hydrology 609 (2022)
  127726.

\bibitem{pradhan2023spatial}
B.~Pradhan, S.~Lee, A.~Dikshit, H.~Kim, Spatial flood susceptibility mapping
  using an explainable artificial intelligence (xai) model, Geoscience
  Frontiers 14~(6) (2023) 101625.

\bibitem{choi2025flo}
H.~Choi, H.~Woo, M.~Kim, H.~Ryu, J.-H. Lee, S.~Lee, S.~J. Noh, Flo-sr: Deep
  learning-based urban flood super-resolution model, Journal of Hydrology 661
  (2025) 133529.
\newblock \href {https://doi.org/10.1016/j.jhydrol.2025.133529}
  {\path{doi:10.1016/j.jhydrol.2025.133529}}.

\bibitem{lowe2021u}
R.~L{\"o}we, J.~B{\"o}hm, D.~G. Jensen, J.~Leandro, S.~H. Rasmussen,
  U-flood--topographic deep learning for predicting urban pluvial flood water
  depth, Journal of Hydrology 603 (2021) 126898.

\bibitem{wang2024time}
Z.~Wang, H.~Lyu, G.~Fu, C.~Zhang, Time-guided convolutional neural networks for
  spatiotemporal urban flood modelling, Journal of Hydrology 645 (2024) 132250.
\newblock \href {https://doi.org/10.1016/j.jhydrol.2024.132250}
  {\path{doi:10.1016/j.jhydrol.2024.132250}}.

\bibitem{herath2025subgrid}
H.~M. V.~V. Herath, L.~Marshall, A.~Saha, S.~Rasnayaka, S.~Seneviratne, Subgrid
  informed neural networks for high-resolution flood mapping, Journal of
  Hydrology 660 (2025) 133329.
\newblock \href {https://doi.org/10.1016/j.jhydrol.2025.133329}
  {\path{doi:10.1016/j.jhydrol.2025.133329}}.

\bibitem{he2023deep}
J.~He, L.~Zhang, T.~Xiao, H.~Wang, H.~Luo, Deep learning enables
  super-resolution hydrodynamic flooding process modeling under
  spatiotemporally varying rainstorms, Water Research 239 (2023) 120057.

\bibitem{zhou2021rapid}
Y.~Zhou, W.~Wu, R.~Nathan, Q.~J. Wang, A rapid flood inundation modelling
  framework using deep learning with spatial reduction and reconstruction,
  Environmental Modelling \& Software 143 (2021) 105112.
\newblock \href {https://doi.org/10.1016/j.envsoft.2021.105112}
  {\path{doi:10.1016/j.envsoft.2021.105112}}.

\bibitem{nearing2024global}
G.~Nearing, D.~Cohen, V.~Dube, M.~Gauch, O.~Gilon, S.~Harrigan, A.~Hassidim,
  D.~Klotz, F.~Kratzert, A.~Metzger, et~al., Global prediction of extreme
  floods in ungauged watersheds, Nature 627~(8004) (2024) 559--563.

\bibitem{cohen2026extending}
D.~Cohen, R.~Amira, R.~Aschner, Y.~Carny, B.~Feinstein, H.~Fester, S.~Fronman,
  M.~Gauch, O.~Gilon, R.~Green, et~al., Extending medium-range global flood
  forecasts: The google global flood forecasting model version 2, EGUsphere
  2026 (2026) 1--31.
\newblock \href {https://doi.org/10.5194/egusphere-2026-2283}
  {\path{doi:10.5194/egusphere-2026-2283}}.

\bibitem{hofmann2021floodgan}
J.~Hofmann, H.~Sch{\"u}ttrumpf, Floodgan: Using deep adversarial learning to
  predict pluvial flooding in real time, Water 13~(16) (2021) 2255.

\bibitem{do2023generalizing}
C.~A. Do~Lago, M.~H. Giacomoni, R.~Bentivoglio, R.~Taormina, M.~N.~G. Junior,
  E.~M. Mendiondo, Generalizing rapid flood predictions to unseen urban
  catchments with conditional generative adversarial networks, Journal of
  Hydrology 618 (2023) 129276.

\bibitem{huang2025high}
L.~Huang, F.~Antolini, A.~Mostafavi, R.~Blessing, M.~Garcia, S.~D. Brody,
  High-resolution flood probability mapping using generative machine learning
  with large-scale synthetic precipitation and inundation data, Computer-Aided
  Civil and Infrastructure Engineering 40~(19) (2025) 2859--2875.
\newblock \href {https://doi.org/10.1111/mice.13490}
  {\path{doi:10.1111/mice.13490}}.

\bibitem{yin2024fast}
Z.~Yin, Y.~Saadati, B.~Hu, A.~S. Leon, M.~H. Amini, D.~McDaniel, Fast
  high-fidelity flood inundation map generation by super-resolution techniques,
  Journal of Hydroinformatics 26~(1) (2024) 319--336.
\newblock \href {https://doi.org/10.2166/hydro.2024.228}
  {\path{doi:10.2166/hydro.2024.228}}.

\bibitem{shen2026attention}
H.~Shen, H.~Hu, C.~Xin, Y.~Yang, Y.~Wu, Z.~Liang, J.~Wen, An attention-based
  super-resolution model for high-accuracy urban pluvial flood forecasting in
  coastal megacities, Journal of Hydrology (2026) 135330.

\bibitem{shao2024data}
P.~Shao, J.~Feng, J.~Lu, P.~Zhang, C.~Zou, Data-driven and knowledge-guided
  denoising diffusion model for flood forecasting, Expert Systems with
  Applications 244 (2024) 122908.
\newblock \href {https://doi.org/10.1016/j.eswa.2023.122908}
  {\path{doi:10.1016/j.eswa.2023.122908}}.

\bibitem{ou2025probabilistic}
Z.~Ou, C.~Nai, B.~Pan, Y.~Zheng, C.~Shen, P.~Jiang, X.~Liu, Q.~Tang, W.~Li,
  M.~Pan, Probabilistic diffusion models advance extreme flood forecasting,
  Geophysical Research Letters 52~(15) (2025) e2025GL115705.
\newblock \href {https://doi.org/10.1029/2025GL115705}
  {\path{doi:10.1029/2025GL115705}}.

\bibitem{neo2026flood}
S.~H. Neo, S.~Seneviratne, H.~M. V. V.~H. Herath, A.~Saha, S.~Rasnayaka, L.~A.
  Marshall, Flood-ldm: Generalizable latent diffusion models for rapid and
  accurate zero-shot high-resolution flood mapping, in: Proceedings of the
  IEEE/CVF Winter Conference on Applications of Computer Vision, 2026, pp.
  8063--8072.

\bibitem{yang2025diffusion}
W.~Yang, H.~Ji, L.~Lonzarich, Y.~Song, C.~Shen, Diffusion-based probabilistic
  modeling for hourly streamflow prediction and assimilation, arXiv preprint
  arXiv:2510.08488 (2025).

\bibitem{chen2023fast}
J.~Chen, Y.~Li, S.~Zhang, Fast prediction of urban flooding water depth based
  on cnn- lstm, Water 15~(7) (2023) 1397.

\bibitem{yang2024rapid}
F.~Yang, W.~Ding, J.~Zhao, L.~Song, D.~Yang, X.~Li, Rapid urban flood
  inundation forecasting using a physics-informed deep learning approach,
  Journal of Hydrology 643 (2024) 131998.
\newblock \href {https://doi.org/10.1016/j.jhydrol.2024.131998}
  {\path{doi:10.1016/j.jhydrol.2024.131998}}.

\bibitem{liao2025accelerating}
Y.~Liao, Z.~Wang, H.~Yu, W.~Gao, Z.~Zeng, X.~Li, C.~Lai, Accelerating urban
  flood inundation simulation under spatio-temporally varying rainstorms using
  convlstm deep learning model, Water Resources Research 61~(8) (2025)
  e2025WR040433.
\newblock \href {https://doi.org/10.1029/2025WR040433}
  {\path{doi:10.1029/2025WR040433}}.

\bibitem{wang2026hybrid}
Y.~Wang, J.~L. Goodall, C.~Kumar, D.~McSpadden, S.~A. Barbosa, B.~Roy,
  A.~Shahabi, N.~Tahvildari, A hybrid cnn-lstm surrogate model for
  hyper-resolution spatiotemporal flood forecasting in norfolk, virginia,
  Journal of Hydrology: Regional Studies 64 (2026) 103234.
\newblock \href {https://doi.org/10.1016/j.ejrh.2026.103234}
  {\path{doi:10.1016/j.ejrh.2026.103234}}.

\bibitem{breiman2001random}
L.~Breiman, Random forests, Machine learning 45~(1) (2001) 5--32.

\bibitem{friedman2001greedy}
J.~H. Friedman, Greedy function approximation: a gradient boosting machine,
  Annals of statistics (2001) 1189--1232.

\bibitem{chen2016xgboost}
T.~Chen, C.~Guestrin, Xgboost: A scalable tree boosting system, in: Proceedings
  of the 22nd acm sigkdd international conference on knowledge discovery and
  data mining, 2016, pp. 785--794.

\bibitem{raissi2019physics}
M.~Raissi, P.~Perdikaris, G.~E. Karniadakis, Physics-informed neural networks:
  A deep learning framework for solving forward and inverse problems involving
  nonlinear partial differential equations, Journal of Computational physics
  378 (2019) 686--707.

\bibitem{li2020fourier}
Z.~Li, N.~Kovachki, K.~Azizzadenesheli, B.~Liu, K.~Bhattacharya, A.~Stuart,
  A.~Anandkumar, Fourier neural operator for parametric partial differential
  equations, arXiv preprint arXiv:2010.08895 (2020).

\bibitem{scarselli2008graph}
F.~Scarselli, M.~Gori, A.~C. Tsoi, M.~Hagenbuchner, G.~Monfardini, The graph
  neural network model, IEEE transactions on neural networks 20~(1) (2009)
  61--80.

\bibitem{kipf2016semi}
T.~N. Kipf, M.~Welling, Semi-supervised classification with graph convolutional
  networks, arXiv preprint arXiv:1609.02907 (2016).

\bibitem{battaglia2018relational}
P.~W. Battaglia, J.~B. Hamrick, V.~Bapst, A.~Sanchez-Gonzalez, V.~Zambaldi,
  M.~Malinowski, A.~Tacchetti, D.~Raposo, A.~Santoro, R.~Faulkner, et~al.,
  Relational inductive biases, deep learning, and graph networks, arXiv
  preprint arXiv:1806.01261 (2018).

\bibitem{elman1990finding}
J.~L. Elman, Finding structure in time, Cognitive science 14~(2) (1990)
  179--211.

\bibitem{hochreiter1997long}
S.~Hochreiter, J.~Schmidhuber, Long short-term memory, Neural computation 9~(8)
  (1997) 1735--1780.

\bibitem{ronneberger2015u}
O.~Ronneberger, P.~Fischer, T.~Brox, U-net: Convolutional networks for
  biomedical image segmentation, in: International Conference on Medical image
  computing and computer-assisted intervention, Springer, 2015, pp. 234--241.

\bibitem{goodfellow2014generative}
I.~J. Goodfellow, J.~Pouget-Abadie, M.~Mirza, B.~Xu, D.~Warde-Farley, S.~Ozair,
  A.~Courville, Y.~Bengio, Generative adversarial nets, Advances in neural
  information processing systems 27 (2014).

\bibitem{salimans2016improved}
T.~Salimans, I.~Goodfellow, W.~Zaremba, V.~Cheung, A.~Radford, X.~Chen,
  Improved techniques for training gans, Advances in neural information
  processing systems 29 (2016).

\bibitem{metz2016unrolled}
L.~Metz, B.~Poole, D.~Pfau, J.~Sohl-Dickstein, Unrolled generative adversarial
  networks, arXiv preprint arXiv:1611.02163 (2016).

\bibitem{arjovsky2017wasserstein}
M.~Arjovsky, S.~Chintala, L.~Bottou, Wasserstein generative adversarial
  networks, in: International conference on machine learning, Pmlr, 2017, pp.
  214--223.

\bibitem{ho2020denoising}
J.~Ho, A.~Jain, P.~Abbeel, Denoising diffusion probabilistic models, Advances
  in neural information processing systems 33 (2020) 6840--6851.

\bibitem{rombach2022high}
R.~Rombach, A.~Blattmann, D.~Lorenz, P.~Esser, B.~Ommer, High-resolution image
  synthesis with latent diffusion models, in: Proceedings of the IEEE/CVF
  conference on computer vision and pattern recognition, 2022, pp.
  10684--10695.

\bibitem{song2023consistency}
Y.~Song, P.~Dhariwal, M.~Chen, I.~Sutskever, Consistency models, in:
  Proceedings of the 40th International Conference on Machine Learning, Vol.
  202 of Proceedings of Machine Learning Research, PMLR, 2023, pp.
  32211--32252.

\bibitem{fukami2019super}
K.~Fukami, K.~Fukagata, K.~Taira, Super-resolution reconstruction of turbulent
  flows with machine learning, Journal of Fluid Mechanics 870 (2019) 106--120.

\bibitem{fukami2021machine}
K.~Fukami, K.~Fukagata, K.~Taira, Machine-learning-based spatio-temporal super
  resolution reconstruction of turbulent flows, Journal of Fluid Mechanics 909
  (2021) A9.

\bibitem{buster2025second}
G.~Buster, B.~N. Benton, D.~Rastogi, S.-C. Kao, G.~Castelao, J.~Eisenman,
  Second-generation downscaled earth system model data using generative machine
  learning, Data in Brief 61 (2025) 111774.
\newblock \href {https://doi.org/10.1016/j.dib.2025.111774}
  {\path{doi:10.1016/j.dib.2025.111774}}.

\bibitem{benton2025super}
B.~N. Benton, G.~Buster, P.~Pinchuk, A.~Glaws, R.~N. King, G.~Maclaurin,
  I.~Chernyakhovskiy, Super-resolution for renewable energy resource data with
  wind from reanalysis data and application to ukraine, Energies 18~(14) (2025)
  3769.

\bibitem{stengel2020adversarial}
K.~Stengel, A.~Glaws, D.~Hettinger, R.~N. King, Adversarial super-resolution of
  climatological wind and solar data, Proceedings of the National Academy of
  Sciences 117~(29) (2020) 16805--16815.

\bibitem{sallam2025tropical}
O.~Sallam, A.~Glaws, M.~Sanchez-Gomez, G.~Deskos, A.~Cortiella, R.~King,
  M.~F{\"u}rth, Tropical cyclone super resolution using conditional diffusion
  denoising probabilistic model from mesoscale simulation to les, Environmental
  Research Communications 7~(12) (2025) 125025.
\newblock \href {https://doi.org/10.1088/2515-7620/ae2a96}
  {\path{doi:10.1088/2515-7620/ae2a96}}.

\bibitem{leutbecher2008ensemble}
M.~Leutbecher, T.~N. Palmer, Ensemble forecasting, Journal of Computational
  Physics 227~(7) (2008) 3515--3539.

\bibitem{price2025probabilistic}
I.~Price, A.~Sanchez-Gonzalez, F.~Alet, T.~R. Andersson, A.~El-Kadi,
  D.~Masters, T.~Ewalds, J.~Stott, S.~Mohamed, P.~Battaglia, R.~Lam,
  M.~Willson, Probabilistic weather forecasting with machine learning, Nature
  637~(8044) (2025) 84--90.

\bibitem{xia2019full}
X.~Xia, Q.~Liang, X.~Ming, A full-scale fluvial flood modelling framework based
  on a high-performance integrated hydrodynamic modelling system (hipims),
  Advances in Water Resources 132 (2019) 103392.

\bibitem{xia2025synxflow}
X.~Xia, X.~Ming, Synxflow: A gpu-accelerated python package for multi-hazard
  simulations, Journal of Open Source Software 10~(111) (2025) 7586.
\newblock \href {https://doi.org/10.21105/joss.07586}
  {\path{doi:10.21105/joss.07586}}.

\bibitem{xia2017efficient}
X.~Xia, Q.~Liang, X.~Ming, J.~Hou, An efficient and stable hydrodynamic model
  with novel source term discretization schemes for overland flow and flood
  simulations, Water resources research 53~(5) (2017) 3730--3759.

\bibitem{xia2018new}
X.~Xia, Q.~Liang, A new efficient implicit scheme for discretising the stiff
  friction terms in the shallow water equations, Advances in water resources
  117 (2018) 87--97.

\bibitem{campos2015numerical}
E.~Campos, J.~Wang, Numerical simulation and analysis of the april 2013 chicago
  floods, Journal of Hydrology 531 (2015) 454--474.

\bibitem{nws2013floodloss}
{National Weather Service},
  \href{https://www.weather.gov/media/water/WY13%20Flood%20Loss%20Summary.pdf}{{United
  States Flood Loss Report -- Water Year 2013}}, Tech. rep., {National Oceanic
  and Atmospheric Administration}, accessed: 2026-05-24 (2013).
\newline\urlprefix\url{https://www.weather.gov/media/water/WY13%20Flood%20Loss%20Summary.pdf}

\bibitem{bonnin2006precipitation}
G.~M. Bonnin, D.~Martin, B.~Lin, T.~Parzybok, M.~Yekta, D.~Riley, {NOAA Atlas
  14, Precipitation-Frequency Atlas of the United States, Volume 2 Version 3.0:
  Ohio River Basin and Surrounding States}, Tech. rep., NOAA, National Weather
  Service, Silver Spring, MD (2006).

\bibitem{salimans_2022}
T.~Salimans, J.~Ho, \href{https://arxiv.org/abs/2202.00512}{Progressive
  distillation for fast sampling of diffusion models}, CoRR abs/2202.00512
  (2022).
\newblock \href {http://arxiv.org/abs/2202.00512} {\path{arXiv:2202.00512}}.
\newline\urlprefix\url{https://arxiv.org/abs/2202.00512}

\bibitem{he2016deep}
K.~He, X.~Zhang, S.~Ren, J.~Sun, Deep residual learning for image recognition,
  in: Proceedings of the IEEE conference on computer vision and pattern
  recognition, 2016, pp. 770--778.

\bibitem{zhang2019self}
H.~Zhang, I.~Goodfellow, D.~Metaxas, A.~Odena, Self-attention generative
  adversarial networks, in: International conference on machine learning, PMLR,
  2019, pp. 7354--7363.

\bibitem{wang2006efficient}
L.~Wang, H.~Liu, An efficient method for identifying and filling surface
  depressions in digital elevation models for hydrologic analysis and
  modelling, International Journal of Geographical Information Science 20~(2)
  (2006) 193--213.

\bibitem{DirkEilander2021}
D.~Eilander, M.~Visser,
  \href{https://doi.org/10.5281/zenodo.5534756}{Deltares/pyflwdir: v0.5}
  (2021).
\newblock \href {https://doi.org/10.5281/zenodo.5534756}
  {\path{doi:10.5281/zenodo.5534756}}.
\newline\urlprefix\url{https://doi.org/10.5281/zenodo.5534756}

\bibitem{strahler1957quantitative}
A.~N. Strahler, Quantitative analysis of watershed geomorphology, Eos,
  Transactions American Geophysical Union 38~(6) (1957) 913--920.

\bibitem{nobre2016hand}
A.~D. Nobre, L.~A. Cuartas, M.~R. Momo, D.~L. Severo, A.~Pinheiro, C.~A. Nobre,
  Hand contour: a new proxy predictor of inundation extent, Hydrological
  Processes 30~(2) (2016) 320--333.

\bibitem{nardi2019gfplain250m}
F.~Nardi, A.~Annis, G.~Di~Baldassarre, E.~Vivoni, S.~Grimaldi, Gfplain250m, a
  global high-resolution dataset of earth’s floodplains, Scientific data 6
  (2019) 180309.
\newblock \href {https://doi.org/10.1038/sdata.2018.309}
  {\path{doi:10.1038/sdata.2018.309}}.

\bibitem{kingma_adam_2014}
D.~P. Kingma, J.~Ba, \href{https://arxiv.org/abs/1412.6980}{Adam: {A} {Method}
  for {Stochastic} {Optimization}}, version Number: 9 (2014).
\newblock \href {https://doi.org/10.48550/ARXIV.1412.6980}
  {\path{doi:10.48550/ARXIV.1412.6980}}.
\newline\urlprefix\url{https://arxiv.org/abs/1412.6980}

\bibitem{matheson1976scoring}
J.~E. Matheson, R.~L. Winkler, Scoring rules for continuous probability
  distributions, Management Science 22~(10) (1976) 1087--1096.

\bibitem{hersbach2000decomposition}
H.~Hersbach, Decomposition of the continuous ranked probability score for
  ensemble prediction systems, Weather and Forecasting 15~(5) (2000) 559--570.

\bibitem{gneiting2007strictly}
T.~Gneiting, A.~E. Raftery, Strictly proper scoring rules, prediction, and
  estimation, Journal of the American Statistical Association 102~(477) (2007)
  359--378.

\bibitem{fortin2014why}
V.~Fortin, M.~Abaza, F.~Anctil, R.~Turcotte, Why should ensemble spread match
  the rmse of the ensemble mean?, Journal of Hydrometeorology 15~(4) (2014)
  1708--1713.

\bibitem{talagrand1997evaluation}
O.~Talagrand, R.~Vautard, B.~Strauss, Evaluation of probabilistic prediction
  systems, Proceedings of the ECMWF Workshop on Predictability (1997) 1--25.

\bibitem{hamill2001interpretation}
T.~M. Hamill, Interpretation of rank histograms for verifying ensemble
  forecasts, Monthly Weather Review 129~(3) (2001) 550--560.

\bibitem{brier1950verification}
G.~W. Brier, Verification of forecasts expressed in terms of probability,
  Monthly Weather Review 78~(1) (1950) 1--3.

\end{thebibliography}

\end{document}